\documentclass[12pt]{article}
\usepackage{amsmath,amsfonts}
\usepackage[nosort]{cite}
\usepackage{graphicx}
\makeatletter\@addtoreset{equation}{section}\makeatother

\def\bC {\mathbb{C}}

\def\bR {\mathbb{R}}

\def\bI {\mathbb{I}}

\newcommand{\be}{\begin{equation}}
\newcommand{\ee}{\end{equation}}
\newcommand{\bea}{\begin{eqnarray}}
\newcommand{\eea}{\end{eqnarray}}
\newcommand{\half}{\frac{1}{2}}

\newcommand{\vev}[1]{{\left< {#1} \right>}}

\newcommand{\eqn}[1]{(\ref{#1})}
\newcommand{\nn}{\nonumber}

\newcommand{\ad}{{\dot a}}
\newcommand{\bd}{{\dot b}}
\newcommand{\cd}{{\dot c}}

\newcommand{\alphad}{{\dot \alpha}}
\newcommand{\betad}{{\dot\beta}}
\newcommand{\gammad}{{\dot \gamma}}

\newcommand{\Tr}{{\rm Tr\,}}

\newcommand{\cD}{{\mathcal D}}

\newcommand{\cG}{{\mathcal G}}

\newcommand{\cN}{{\mathcal N}}

\newcommand{\cQ}{{\mathcal Q}}
\newcommand{\cR}{{\mathcal R}}
\newcommand{\cO}{{\mathcal O}}
\newcommand{\cS}{{\mathcal S}}

\newcommand{\cX}{{\mathcal X}}
\newcommand{\cY}{{\mathcal Y}}
\newcommand{\cZ}{{\mathcal Z}}

\renewcommand{\title}[1]{\vbox{\center\LARGE{#1}}\vspace{5mm}}
\renewcommand{\author}[1]{\vbox{\center#1}\vspace{5mm}}
\newcommand{\address}[1]{\vbox{\center\em#1}}
\newcommand{\email}[1]{\vbox{\center\tt#1}\vspace{5mm}}

\begin{document}
\begin{titlepage}
\begin{center}
\vspace{5mm}
\hfill {\tt HU-EP-09/01}\\
\vspace{20mm}
\title{Superprotected $n$-point correlation functions\\ 
of local operators in $\cN=4$ super Yang-Mills}

\author{\large Nadav Drukker and Jan Plefka}
\address{Institut f\"ur Physik, Humboldt-Universit\"at zu Berlin,\\
Newtonstra{\ss}e 15, D-12489 Berlin, Germany}

\email{drukker, plefka@physik.hu-berlin.de}

\end{center}

\abstract{
\noindent
In this paper we study the $n$-point correlation functions of two different families of 
local gauge invariant operators in $\cN=4$ supersymmetric Yang-Mills theory. 
The main idea is to consider the correlation functions of operators which 
all share a number of supersymmetries irrespective of their relative locations.
We achieve this by equipping the operators with explicit space-time dependence. 
We provide evidence by different methods that 
these $n$-point correlators do not receive quantum corrections in 
perturbation theory and are hence given exactly by their tree-level result. 
The arguments rely on explicit checks for 
general four-point correlators, some five-point and six-point correlators and a more 
abstract calculation based on a novel topological twisting of 
$\cN=4$ supersymmetric Yang-Mills theory.
}

\vfill

\end{titlepage}

{\addtolength{\parskip}{-1ex}\tableofcontents}

\section{Introduction}

Great progress has been achieved in the past few years in precision studies of 
$\cN=4$ supersymmetric Yang-Mills (SYM) theory and of the dual string theory 
on $AdS_5\times S^5$ \cite{Maldacena:1997re,GKP,Witten}. 
The problem of finding the exact anomalous scaling dimensions of local operators
has been recast into that of diagonalizing a long-range spin chain model, 
which---assuming integrability---can be solved for asymptotically long 
operators by the Bethe ansatz \cite{Minahan:2002ve, Beisert:2003tq, Bena:2003wd, 
Betheeq,Beisert:2006ez}\footnote{For reviews see \cite{intrevs}.}. 
The most obvious remaining problem is the understanding of wrapping interactions, 
which affect short operators at lower loop orders \cite{wrapping,Janik}.

Beyond that, one would want to go over and obtain all loop results for three-point 
correlation functions and 
more generally $n$-point correlators of local gauge invariant operators. 
In the case of three-point correlators, they are well understood when all three operators 
are chiral primaries (1/2 BPS operators) still from the early days of the $AdS$/CFT correspondence \cite{Lee:1998bxa}. These three-point functions 
are protected from radiative corrections and are given precisely 
by the free field theory approximation. 
The case of four-point functions is much more complicated, as they are subject 
to quantum corrections 
\cite{D'Hoker:1999pj, D'Hoker:1999ni, Eden:1999kh, Dolan:2000ut, 
Heslop:2002hp, Arutyunov:2002fh, Arutyunov:2003ae, Arutyunov:2003ad},
while very little is known about higher-point functions.

The lack of quantum corrections to the two-point and three-point functions of chiral 
primary operators can be attributed to the fact that all the operators in the correlation 
function share a number of common supersymmetries. 
A single operator is annihilated by 24 supercharges: When the operator is 
located the origin, $x^\mu=0$, these are all of the superconformal generators 
(denoted as $S$) and half of the Poincar\'e supercharges (denoted as $Q$). 
At other space-time positions these are 24 other combinations of 
these supercharges. The most general combination of three operators 
of this type at arbitrary space-time positions
will still preserve eight supercharges,%
\footnote{Note that they may break all the $Q$'s and preserve only $S$'s, 
which is not considered a supersymmetric configuration, but by our 
counting it would be.} 
since each breaks only eight. Four operators, on the other hand, 
will generically not share any supersymmetries, which is exactly when 
radiative corrections start to occur.

The object of this paper is to find families of operators which share more 
supercharges than generic $1/2$ BPS operators. One may hope that the correlator 
of four or more such operators, who share a number of supercharges, will be 
simpler than that of $n$-point correllators of generic 1/2 BPS operators. 
This is indeed true in the two examples of families of operators we present. 

A trivial example is the case of operators all preserving the same 
super-Poincar\'e generators. If we consider one of the complex scalar fields 
of the $\cN=4$ supersymmetry multiplet $Z=\Phi^5+i\Phi^6$ and build operators 
out of it, then
\be
\vev{\Tr Z^{J_1}(x_1)\ \Tr Z^{J_2}(x_2)\ \cdots\ \Tr Z^{J_n}(x_n)}=0\,.
\ee
This is obvious since they all carry positive charge under a $U(1)$ 
subgroup of the $R$-symmetry group. But a similar statement is 
{\em almost} true also if there was only $\cN=1$ supersymmetry and no 
$R$-charge. In that case chiral primary operators form a ring and do not 
interact with each other. Their classical $n$-point function vanishes and 
they only receive divergent quantum corrections due to instantons. Our 
examples will share many features with these chiral rings.

In the other examples we present in this paper, the choice of operator 
is dependent on its spatial position. At different locations the operators 
will be made of different linear combinations of the scalar fields.

The way we realize this is by taking local operators of the form
\be
\label{1.2}
\Tr [u_I(x)\, \Phi^I(x)]^J \, ,
\ee
with $u_I(x)$ complex six-vectors. These operators are $1/2$ BPS if 
$u_I(x)\, u_I(x)=0$ and furthermore, suitable choices of the $u^I(x)$ gives 
operators that share some conserved supercharges irrespective of the 
position $x^\mu$ in some submanifold of space. 
In the following two sections we give two examples of such constructions.
The first example in Section~\ref{sec-R4} allows the operators to be at arbitrary 
points $x^\mu\in\bR^4$ and they involve all six real scalars of $\cN=4$ SYM. 
The second example turns on only three of the scalars and the operators 
are restricted to $x^\mu\in\bR^2$ in space-time.

We study the operators in a variety of ways. After presenting each 
example we show the supercharges that are preserved by the relevant 
operators. We then 
study how the symmetry generators of $PSU(2,2|4)$ act on the operators. 
In both cases there are linear combinations of symmetry generators whose 
action on the operators is particularly simple, these generators arise 
naturally in topologically twisted versions of $\cN=4$ SYM. In the 
examples we consider the topological twisting involves conformal generators 
and not merely the Poincar\'e group. We will not study the topological 
twistings in detail, but we expect that a lot of the features that we 
point out can be proven by use of topological gauge theories.

We then concentrate on perturbative calculations of specific $n$-point functions 
of the operators we constructed. 
Using previously found results for the four-point function of generic 
chiral primary operators we can immediately show that for our operators 
there are no perturbative corrections. In a companion paper \cite{us} we 
develop a simple formula for the one-loop correction to all $n$-point 
functions of chiral primary 
operators. In that paper we use this formula to evaluate some five-point 
functions and a six-point function at one loop. Here we show that when 
concentrating on our special operators, these one-loop quantum 
corrections vanish.

In the next two sections we study the details of the two constructions, 
relegating more technical details of the supersymmetry algebra 
to appendices. We conclude in Section~\ref{sec-discussion} with a 
summary of our results and an extensive discussion of possible 
generalizations and uses of these ideas.

\section{Example I: $1/16$ BPS $n$-point functions on $\bR^4$}
\label{sec-R4}

For our first example we take the six real scalars of $\cN=4$ SYM theory 
$\Phi^1,\,\ldots ,\Phi^6$ and at an arbitrary point $x^\mu\in\bR^4$ define 
the field
\begin{equation}
C(x)=2ix^\mu\Phi^\mu(x)
+i\left(1-(x^\mu)^2\,\right)\Phi^5(x)
+\left(1+(x^\mu)^2\,\right)\Phi^6(x)\, ,
\label{C}
\end{equation}
note that this corresponds to the six-vector in \eqn{1.2} 
\be
\label{uIC}
u_I(x)=\Big(2ix^1\,,2ix^2\,,2ix^3\,,2ix^4\,,i(1-(x^\mu)^2)\,,1+(x^\mu)^2\Big)\,,
\ee
which indeed satisfies $u(x)^2=0$. 
Using $C(x)$ we can then build $1/2$ BPS gauge invariant local operators
\be
\Tr\,C(x)^J\,.
\ee
In the definition of $C$ we assigned to four of the six scalars a Lorentz index
$\mu$, which is the first indication that some topological twisting is involved 
in the construction. Note that the different terms appearing in the 
definition have varying scaling dimensions,
which could be fixed by adding appropriate powers of an arbitrary length-scale.
For simplicity we set this dimensionful constant to unity.
The field $C$ was considered in the past in \cite{deMedeiros:2001kx,Pestun}, 
for somewhat different motivations. We present our point of view on these 
operators and will rely on some of the results of \cite{deMedeiros:2001kx}
below.

When considering the gauge theory on $S^4$ these operators can also be 
written in a compact form. Representing the sphere in flat $\bR^5$ we 
have
\begin{equation}
C(x)=i\Phi^m(x) x^m+\Phi^6(x)\,,\qquad
m=1,\,\cdots,5\,,\qquad
(x^m)^2=1\,.
\end{equation}

We may also write the sphere as the base of the light-cone in $\bR^{5,1}$ and
now
\begin{equation}
C(x)\propto x^i\Phi^i(x)\,,
\label{light-cone}
\end{equation}
with $i=1,\,\ldots,6$ and in the sixth direction a $(-i)$ is included.

\subsection{Supersymmetry}
\label{sec-R4-susy}

We wish to calculate now the supercharges that are preserved by the 
field $C$ at an arbitrary point in space. A compact way of writing the 
general variation of a scalar $\Phi^i$ under both the Poincar\'e and 
conformal supercharges is as
\be
\delta\Phi^i=\bar\psi \rho^i\, \gamma^5\,\epsilon\,,\qquad
\epsilon=\epsilon_0+\gamma_\mu x^\mu\epsilon_1\,.
\ee
Here $\psi$ is the gluino which transforms in the spinor representation of the 
Lorentz and $SO(6)$ R-symmetry groups, $\rho^i$ are the $SO(6)$ 
gamma matrices, while $\gamma_\mu$ are those of the spatial $SO(4)$ 
and we take them to commute with each-other. $\epsilon_0$ and 
$\epsilon_1$ are constant 16-component spinors which are the parameters for 
the super-Poincar\'e and superconformal transformations respectively. 
Our notations and details of the superconformal algebra are listed in 
Appendix~\ref{app-notation}.

Applying this to our local field $C(x)$ of \eqn{C} gives
\begin{equation}
\delta C(x)
=\bar\psi\left(2ix^\mu\rho^\mu\gamma^5
+i(1-(x_\mu)^2)\rho^5\gamma^5+(1+(x_\mu)^2)\rho^6\gamma^5\right)
(\epsilon_0+\gamma_\mu x^\mu\epsilon_1)\,.
\end{equation}
Expanding and separating into terms with different $x$ dependences gives
among others, the equations
\begin{equation}
(\rho^6+i\rho^5)\epsilon_0=0\,,\qquad
(\rho^6-i\rho^5)\epsilon_1=0\,,\qquad
i\rho^\mu\epsilon_0+\rho^6\gamma^\mu\epsilon_1=0\,.
\label{susy-var1}
\end{equation}
All the other equations are automatically solved once we impose
these conditions, which are also not independent. The first two
are a consequence of the last ones, which can be rewritten as
\begin{equation}
\gamma^1\rho^1\epsilon_0=
\gamma^2\rho^2\epsilon_0=
\gamma^3\rho^3\epsilon_0=
\gamma^4\rho^4\epsilon_0=
i\rho^6\epsilon_1\,.
\end{equation}
Since $\epsilon_0$ and $\epsilon_1$ arise from chiral spinors in 10-dimensions,%
\footnote{In our conventions 
$\Gamma^{10}\epsilon_0=\epsilon_0$ and $\Gamma^{10}\epsilon_1=-\epsilon_1$ 
with $\Gamma^{10}=i\gamma^1\gamma^2\gamma^3\gamma^4
\rho^1\rho^2\rho^3\rho^4\rho^5\rho^6$.} this
automatically sets the correct relation between the last two matrices
$\rho^5$ and $\rho^6$ acting on it, just as the first equation in
(\ref{susy-var1}). Then $\epsilon_1$ is completely defined in
terms on $\epsilon_0$.

The above conditions on $\epsilon_0$ can be rearranged as
\begin{equation}
\gamma^{\mu\nu}\epsilon_0=-\rho^{\mu\nu}\epsilon_0\,,\qquad
\mu,\,\nu=1,\cdots,4\,.
\label{LorRrel}
\end{equation}
Now note that $\gamma^{\mu\nu}$ are the generators of the Lorentz
group in the spinor representation while $\rho^{\mu\nu}$ are
six out of the 15 generators of the $R$-symmetry group, also
in a spinor representation. This equation suggests taking the
diagonal sum of the two groups and imposing that $\epsilon_0$
is a singlet under the diagonal group.

$\epsilon_0$ is the sum of a chiral spinor
$\epsilon_{0A}^{+\,\alpha}$ transforming in the
$({\bf2},{\bf1},{\bf4})$ representation of
$SU(2)_L\times SU(2)_R\times SU(4)$ and an anti-chiral spinor
$\epsilon_0^{-\,\alphad A}$ in the
$({\bf1},{\bf2},\bar{\bf 4})$ representation.
The above equation
suggests to break the $R$-symmetry also to $SU(2)_A\times SU(2)_B$,
such that the spinor is decomposed as
$\bf4\to ({\bf2},{\bf1})\oplus({\bf1},{\bf2})$. We will use dotted
lowercase roman indices for $SU(2)_A$ and undotted ones for $SU(2)_B$.

Under this decomposition the most general supercharge is generated by
\begin{equation}
\epsilon_{0\,a}^{+\,\alpha}Q_\alpha^a
+\dot\epsilon_{0}^{+\,\alpha\ad}\dot Q_{\alpha\ad}
-\epsilon_{1\,\alphad a}^{-}\bar S^{\alphad a}
-\dot\epsilon_{1\,\alphad}^{-\,\ad}\dot{\bar S}^\alphad_\ad
+\epsilon_{1\,\alpha}^{+\,a}S^\alpha_a
+\epsilon_{0}^{-\,\alphad a}\bar Q_{\alphad a}
-\dot\epsilon_{1\,\alpha\ad}^{+}\dot S^{\alpha\ad}
-\dot\epsilon_{0\,\ad}^{-\,\alphad}\dot{\bar Q}_\alphad^\ad\, .
\label{general-susy-R4}
\end{equation}
Details are given in Appendix~\ref{app-R4-susy}.

We may now view the above equation \eqn{LorRrel}
as relating $SU(2)_L$ with
$SU(2)_B$ and $SU(2)_R$ with $SU(2)_A$, so we need to consider
only the spinors with either both dotted or both undotted
space-time and $R$-symmetry indices. Furthermore, the requirement
that they are a singlet of the diagonal group means that they
can be written as
\begin{equation}
\epsilon_{0a}^{+\,\alpha}=\delta_a^\alpha\,\epsilon_0^+\,,\qquad
\dot{\epsilon}_{0\ad}^{-\,\alphad}
=\delta^\alphad_{\ad}\,\dot\epsilon_0^-\,,
\end{equation}
where $\epsilon_0^-$ and $\dot\epsilon_0^-$ will serve as the
two parameters of the unbroken supersymmetries.

$\epsilon_1$ can now be determined through the equation
$i\rho^6\epsilon_1=\gamma^1\rho^1\epsilon_0$. The generator
$\rho^{16}$ changes a dotted index into an undotted one, as
does the single gamma matrix $\gamma^1$.
In our notations in Appendix~\ref{app-notation} the gamma matrix with 
lower indices is $\gamma^1_{\alphad\alpha}=i\tau^1$ and in 
Appendix~\ref{app-R4-susy} one finds that
$(\rho^{51}+i\rho^{61})^{\,a\ad}=-i\tau^1$, so
\begin{equation}
\epsilon_{1\alpha}^{+\,a}
=\tau^1_{\alpha\alphad}\tau^{1\,a\ad}
\dot{\epsilon}_{0\ad}^{-\,\alphad}
=\delta^a_\alpha \dot{\epsilon_0}^-\,,\qquad
\dot\epsilon^{-\,\ad}_{1\alphad}
=\tau^1_{\alpha\alphad}\tau^{1\,a\ad}
\epsilon_{0a}^{+\,\alpha}
=\delta^\ad_{\alphad}\,\epsilon_0^+\,.
\end{equation}
Plugging this into \eqn{general-susy-R4} we find that the supercharges 
that annihilate all of the operators $C$, regardless of their positions, are
\begin{equation}
\cQ^+=\delta^\alpha_a\,Q_\alpha^a
-\delta^\ad_\alphad\,\dot{\bar S}_\ad^{\alphad}\,,\qquad
\cQ^-=\delta_\ad^{\alphad}\,\dot{\bar Q}^\ad_{\alphad}
-\delta^a_\alpha\,S^\alpha_a\,.
\end{equation}

While $C$ at a specific position preserves 24 supercharges, like any 
other chiral field, the fields $C$ all share two supercharges irrespective 
of their positions. In special cases, when the positions are not 
totally generic there will be enhanced supersymmetry:
\begin{itemize}
\item
Clearly at two different points $C(x_1)$ and $C(x_2)$ share sixteen supercharges. 

\item
At three different points operators built out of $C(x_i)$ share only eight 
supercharges, which is the same as for generic 
three $1/2$ BPS local operators. Furthermore, any three operators define a line 
or a circle on $\bR^4$. If we consider {\em any} number of operators made of 
the $C$s at arbitrary points along the line/circle they do not break any 
more of the supersymmetries and still preserve $1/4$ of the supercharges. 

\item
Likewise considering $C$ at four points, or at any number 
of points on an $S^2$ or an $\bR^2$ subspace, will lead to four preserved 
supercharges.

\item
Five different operators at generic positions are already the 
general case and preserve only two supercharges.
\end{itemize}

\subsection{Twisted symmetry}
\label{sec-R4-twist}
We have seen that operators built out of the field $C$ are all invariant 
under two supercharges $\cQ^\pm$. Here we address how they transform 
under the remaining symmetry generators.

Some of the symmetry involved in the construction of $C$ is apparent already 
on a quick inspection of \eqn{C}. We assigned to four of the scalar fields 
Lorentz indices on $\bR^4$, or in the construction based on the light cone 
\eqn{light-cone}, we assigned a Lorentz index to all six. This suggests that 
$C$ will transform covariantly when combining $R$-symmetry rotations and 
Poincar\'e and conformal transformations.

Indeed in Section~\ref{sec-R4-susy} we saw that the supercharges that 
annihilate $C$ are singlets of a diagonal subgroup of the $SO(5,1)$ 
conformal group and the $SO(6)$ $R$-symmetry group.%
\footnote{These are not the same groups, of course, but both are certain
real subgroups of $SL(4,\bC)$. We are working mostly at the level of the 
algebra and are therefore not affected much by this. A more careful 
treatments is given in \cite{deMedeiros:2001kx} where it is argued that 
the $R$-symmetry group should really be also $SO(5,1)$.
}
A simple way of finding the twisted symmetry is to take the anti-commutators 
of $\cQ^\pm$ with the other supercharges. As is shown in 
Appendix~\ref{app-R4-susy}, this leads to the combinations of 
bosonic symmetries \eqn{twisted-comm-R4}
\begin{equation}
\begin{aligned}
\hat P_\mu&=P_\mu+R_{5\mu}+iR_{6\mu}\,,\\
\hat J_{\mu\nu}&=J_{\mu\nu}+R_{\mu\nu}\,,\\
\hat D&=D+iR_{56}\,,\\
\hat K_\mu&=K_\mu+R_{5\mu}-iR_{6\mu}\,.
\label{R4-twisted}
\end{aligned}
\end{equation}

Our construction therefore involves an identification of the $R$-symmetry
group and the space-time group, which is the way one obtains topological
theories out of theories with extended supersymmetries. Usually these
constructions twist an $SU(2)$ in space-time by an $SU(2)$ $R$-symmetry.
Here the twist involves also the conformal generators and as we shall see 
our other example in Section~\ref{sec-R2} 
is also associated to topological twistings of 
a subgroup of the conformal group.

In \eqn{symmetry}, \eqn{R-symmetry} 
the action of the bosonic symmetry generators on scalar fields
is written out. From that we can derive the action of the combined 
generators in \eqn{R4-twisted} on our field
$C$, incorporating the explicit space-time dependence
\begin{equation}
\begin{aligned}
\hat P_\mu\,C&=\partial_\mu C\,,\\
\hat J_{\mu\nu}\,C
&=(x_\mu\partial_\nu-x_\nu\partial_\mu)C\,,\\
\hat D\,C&=x^\mu\partial_\mu C\,,\\
\hat K_\mu\,C &=(2x_\mu x^\nu\partial_\nu-x^2\partial_\mu)C\,.
\end{aligned}
\label{twisted-C}
\end{equation}
Therefore $C$ transforms as a dimension-zero scalar of this
twisted conformal group. Indeed its tree-level two-point function
is given by
\be
\vev{C(x_1)\, C(x_2)}_0= \frac{u_I(x_1)\cdot u_I(x_2)}{(2\pi)^2\, (x_1-x_2)^2}
=\frac{1}{2\pi^2}\, ,
\ee
suppressing the gauge group indices.

The fact that the symmetry generators arise as anti-commutators with 
$\cQ^\pm$ \eqn{twisted-comm-R4} allows us to prove that the $n$-point 
function is position independent. Consider the correlator
\be
\vev{\Tr\,C^{J_1}(x_1)\ \Tr\,C^{J_2}(x_2)\ \cdots\ \Tr\,C^{J_n}(x_n)}\,.
\ee
We use the fact that $\hat P^\mu=\big\{\cQ^+,\,Q_\mu\big\}$, that 
$\cQ^+$ annihilates all $C$'s and the Ward-Takahashi identity associated 
to the symmetry generator $\cQ^+$ to derive
\be
\begin{aligned}
&\frac{\partial}{\partial\,x_1^\mu}
\vev{\Tr\,C^{J_1}(x_1)\ \Tr\,C^{J_2}(x_2)\ \cdots\ \Tr\,C^{J_n}(x_n)}
\\&\hskip1cm
=\cQ^+\vev{J_1\Tr\left[\big\{Q_\mu,\,C\big\}C^{J_1-1}(x_1)\right]\ 
\Tr\,C^{J_2}(x_2)\ \cdots\ \Tr\,C^{J_n}(x_n)}
=0\,.
\end{aligned}
\ee
This statement is exact regardless of any quantum corrections (including 
non-perturbative ones).

Furthermore, it was proven in \cite{deMedeiros:2001kx} that the action of 
$\cN=4$ SYM theory when restricted to the zero instanton sector is 
$\cQ^\pm$--exact, {\em i.e.} $\cS_\text{pert}=\big\{\cQ^\pm,\,\Psi^\pm\big\}$ 
with some $\Psi^\pm$. This implies that the $n$-point function receives 
no perturbative corrections%
\footnote{This is true when suitably normalizing the operators  
to absorb the powers of $g_{YM}$ coming from the free propagators.}
\be
\begin{aligned}
&\frac{\partial}{\partial\,g_{YM}^2}
\vev{\Tr\,C^{J_1}(x_1)\ \Tr\,C^{J_2}(x_2)\ \cdots\ 
\Tr\,C^{J_n}(x_n)}_\text{pert}
\\&\hskip2cm
\propto\cQ^+\vev{\Psi^\pm\ \Tr\,C^{J_1}(x_1)\ \Tr\,C^{J_2}(x_2)\ \cdots\ 
\Tr\,C^{J_n}(x_n)}_\text{pert}
=0\, .
\label{superprotectedC}
\end{aligned}
\ee
These results are very reminiscent of those for operators in the chiral 
ring of theories with $\cN=1$ supersymmetry. There one can further use 
cluster decomposition to prove that the $n$-point function vanishes 
perturbatively and receives contributions only from the 
Veneziano-Yankielowicz superpotential.

In our case the theory is conformal, so there is no cluster decomposition. 
The $n$-point function is not zero perturbatively, but given by tree-level 
contractions, as is discussed in the next subsection. We have not 
evaluated the instanton corrections.

In addition to the two supercharges annihilating the field $C$, and the 
fifteen symmetry generators that act on it covariantly \eqn{R4-twisted}, there 
are also fifteen more fermionic generators under which it transforms 
covariantly \eqn{twisted-Qs}. They are given by the sum of the two off-diagonal 
blocks in \eqn{broken-matrix}. Together with the bosonic generators 
\eqn{R4-twisted} they form the superalgebra $Q(4)$.

\subsection{Explicit perturbative calculations}
\label{PertI}

As argued already in the last section, the $n$-point functions of operators 
made of powers of $C$
\be
\label{npointC}
\Bigl \langle \Tr C^{J_1}(x_1)\, \Tr C^{J_2}(x_2)
\ldots \Tr C^{J_n}(x_n) \Bigr \rangle,
\ee
receive no radiative corrections in perturbation theory and may thus be called 
``superprotected''. This is a property known to be true for two-pint and three-point
functions of {\em all} chiral primary operators, the novelty here is that it extends
to $n$-point functions of the special chiral primary operators made of the field 
$C$.

The argument given in the preceding section for the vanishing of all perturbative 
corrections to the $n$-point function is based on the proof of 
\cite{deMedeiros:2001kx} that the action is $Q^\pm$ exact. 
We want to back up this elegant formal argument through explicit computations  
of the first quantum correction to all $n$-point functions and all the perturbative 
corrections to the four-point function of these operators (all with the same $J$). 
These considerations will also be of later use in Section~\ref{sec-R2}.

In \cite{us} we derive a compact expression for the planar 
one-loop quantum correction to all $n$-point functions of operators of the form
\begin{equation}
\label{Okdef}
{\cal O}^u_J (x) =  \Tr\big[u_I\,\Phi^{I}(x)\big]^J\,,
\end{equation}
where the $u_I$ are arbitrary complex six-component vectors obeying 
$u_I\, u_I=0$. This makes ${\cal O}^u_J$ a chiral primary.

The one-loop correction to the $n$-point function is written as a sum 
over all possible choices of four of the operators, with labels $i$, $j$, $k$ 
and $l$. One field from each of these operators interacts through 
a combined four-point vertex $D_{ijkl}$ and the rest of the fields of 
these four operators have to be contracted with all the other operators 
in a planar way (on a disc, with these four operators on the boundary). 
This can be written as
\be
\vev{\cO^{u_1}_{J_1}\cdots\cO^{u_n}_{J_n}}_\text{1-loop}
=\sum_{i,j,k,l}J_iJ_jJ_kJ_l\,D_{ijkl} 
\,\vev{\cO^{u_i}_{J_i-1}\cO^{u_j}_{J_j-1}\cO^{u_k}_{J_k-1}\cO^{u_l}_{J_l-1}\,\Big|\,
\prod_{p\neq i,j,k,l}\cO^{u_p}_{J_p}}_\text{tree, disc}
\label{4-pt}
\ee
The effective interaction vertex $D$ is%
\footnote{For clarity we sometimes replace the general indices $ijkl$ 
 with $1234$.}
\be
\label{D-graph}
D_{1234}=
\frac{\lambda}{32\pi^2}\,\Phi(s,t)\big(2\,[13][24]+(s-1-t)[14][23]+(t-1-s)\,[12][34]\big),
\ee
where $[ij]$ are the tree level contractions (without gauge-group indices), while 
$s$ and $t$ are the cross-ratios
\be
\label{cross-ratios-R4}
[ij]\equiv \frac{1}{(2\pi)^2}\,\frac{u_I^i\cdot u_I^j}{x_{ij}^2}\,,\qquad
s=\frac{x_{12}^2\, x_{34}^2}{x_{13}^2\, x_{24}^2}\, , \qquad
t=\frac{x_{14}^2\, x_{23}^2}{x_{13}^2\, x_{24}^2}\, ,\qquad
x_{ij}\equiv x_i-x_j\,,
\ee
and $\Phi(s,t)$ is the scalar box integral \cite{David}
\be
\Phi(s,t)=\frac{x_{13}^2\,x_{24}^2}{\pi^2}
\int d^4 x_5 \,\frac{1}{x_{15}^2\,x_{25}^2\,x_{35}^2\,x_{45}^2}\,.
\label{Phi}
\ee
One last thing to note, in equation \eqn{4-pt} one should sum over three 
inequivalent orders of the operators: $ijkl$, $ikjl$ and $iklj$, since the tree 
level disc amplitudes with these orderings are generically different. 
It makes some sense to combine all these terms together, since 
$D_{1234}+D_{1324}+D_{1243}=0$, which allows to simplify some 
expressions, but this is not 
necessary for the current calculation. It will be important in Section~\ref{sec-R2}.

With this result it is easy to prove that there are no one-loop corrections to the $n$-point 
function of operators made of the field $C$. In this case we have \eqn{uIC} that 
$u_I^i=u_I(x_i)$ depends on the position $x_i$.
This gives the inner product $u^i_I\cdot u^j_I=2x_{ij}^2$ and hence the 
free-field contractions are all constant
\begin{equation}
[ij]=\frac{1}{2\pi^2}\,.
\end{equation}
Plugging into \eqn{D-graph} we find that $D_{ijkl}=0$, so there are no one-loop 
corrections to {\em any} of the $n$-point functions of our operators.

In the case of four-point functions, we can extend this to an {\em all}-loop statement, 
relying on the results of Arutyunov, Dolan, Osborn and Sokatchev 
\cite{Arutyunov:2002fh,Arutyunov:2003ae}.

Based on superconformal symmetry and additional dynamical input 
these authors showed that the all-loop quantum corrections to the four-point 
amplitude of general chiral primaries of weight $J$ are of a factorized, universal form
\be
\label{R1}
\vev{{\cal O}^{u_1}_J(x_1)\, {\cal O}^{u_2}_J(x_2)\, {\cal O}^{u_3}_J(x_3)\, 
{\cal O}^{u_4}_J(x_4)}_{\text{quant}}
= \mathcal{R}(s,t;\cX,\cY,\cZ)\, {\cal F}_J(s,t;\cX,\cY,\cZ;\lambda)\,,
\ee
where $\cX$, $\cY$ and $\cZ$ are the pair-wise contractions
\begin{equation}
\label{XYZ}
\mathcal{X}=[12][34]\, , \qquad
\mathcal{Y}=[13][24]\, , \qquad
\mathcal{Z}=[14][23]\,.
\end{equation}
The important ingredient in \eqn{R1} is $\cR$, the  
{\em universal polynomial prefactor} which is independent of $J$ or $\lambda$. 
It is given by the simple combination
\be
\begin{aligned}
\label{cRI}
\mathcal{R}&=s\, (\cY-\cX)(\cZ-\cX) + t\, (\cZ-\cX) (\cZ-\cY) +
(\cY-\cX)(\cY-\cZ)
\\&=
\frac{16}{\lambda\,\Phi(s,t)}\left(\cY D_{1234}+\cX D_{1324}+\cZ D_{1243}\right).
\end{aligned}
\ee
Moreover the functions $\mathcal{F}_J$ are known up to two-loop order for $J\leq 4$
\cite{Arutyunov:2003ad}.

Clearly in our case
\begin{equation}
\mathcal{X}=\mathcal{Y}=\mathcal{Z}=\frac{1}{4\pi^4}\,,
\end{equation}
so $\cR=0$ and therefore there are no radiative corrections to the four-point functions.

\section{Example II: $1/8$ BPS $n$-point functions on $\bR^2$}
\label{sec-R2}

We now turn to the discussion of our second example for a superprotected operator.
If we restrict the operator $C$ from Section~\ref{sec-R4} 
to the $(x^1,x^2)$ plane (i.e.~$x^3=x^4=0$) it is
\begin{equation}
C=2ix^1\Phi^1+2ix^2\Phi^2+i(1-(x^\mu)^2)\Phi^5+(1+(x^\mu)^2)\Phi^6\,.
\label{restricted-C}
\end{equation}
These operators will share four supercharges, twice as many as the
most general operators on $\bR^4$. We present in this section another
construction of local operators on this plane made out of only
three of the scalars $\Phi^1$, $\Phi^2$ and $\Phi^3$, which will
also share four supercharges.

Using the complex coordinates $w=x^1+ix^2$ and $\bar w=x^1-ix^2$
define
\begin{equation}
Z=i(1-\bar w^2)\Phi^1+(1+\bar w^2)\Phi^2-2i\bar w\Phi^3\,,
\label{Z}
\end{equation}
which corresponds to the choice in \eqn{1.2}
\be
u_I(\bar w)=\Big(i(1-\bar w^2)\,,1+\bar w^2\,,-2i\bar w\,,0\,,0\,,0\Big).
\label{uI-R2}
\ee
As before we use $Z$ to construct gauge invariant local operators
\be
\Tr\,Z^J(w,\bar w)\,,
\ee
at arbitrary positions on $\bR^2$.

While the definition of $Z$ is different from the restriction of $C$ to 
generic points on $\bR^2$, if we restrict both to a line, they are 
the same up to the choice of scalar fields. For real $w$, for example, 
$Z$ in \eqn{restricted-C} is the same as $C$ \eqn{C} with 
$(\Phi^1,\,\Phi^2,\,\Phi^3)\to(\Phi^5,\,\Phi^6,\,-\Phi^1)$. Indeed 
while generically all $C$s share four supercharges, along a line 
or a circle they share eight.

A nice realization of the same operators $Z$ shows up when considering 
three scalar fields on $S^2$. Using the indices $i,j,k=1,2,3$ both for 
unit three-vectors and for the three scalars, we may define the following 
scalar field
\begin{equation}
Z^i=
(\delta^{ij}-x^ix^j)\Phi^j+i\varepsilon_{ijk}x^j\Phi^k\,.
\label{Z-on-S^2}
\end{equation}
We study operators built out of this field in Appendix~\ref{app-S2}, 
where we explain the spurious superscript in $Z^i$ and how it is related 
to $Z$ in \eqn{Z}.

\subsection{Supersymmetry}
\label{sec-R2-susy}

Examining the invariance of these operators under supersymmetry leads to 
the equations
\begin{equation}
\left(\rho^- -\bar{w}\rho^3 -\bar{w}^2\rho^+\right)
\left(\epsilon_0+(w\gamma^-+\bar{w}\gamma^+)\epsilon_1\right)=0\,,
\end{equation}
where we defined  $\rho^\pm=(\rho^1\pm i\rho^2)/2$ and
$\gamma^\pm=(\gamma^1\pm i\gamma^2)/2$.

Requiring that this is satisfied for all $w$ and $\bar{w}$ leads to
the independent equations
\begin{equation}
\rho^3\epsilon_0-\rho^-\gamma^+\epsilon_1=0\,,\qquad
\rho^+\epsilon_0+\rho^3\gamma^+\epsilon_1=0\,,\qquad
\gamma^-\epsilon_1=0\,.
\label{Zplane-susy}
\end{equation}
We can isolate the following conditions on $\epsilon_1$
\begin{equation}
\gamma^-\epsilon_1=\rho^3\rho^+\epsilon_1=0\,.
\end{equation}
As in Section~\ref{sec-R4-susy}, it proves useful again to consider the 
breaking of the $R$-symmetry group $SO(6)\to SU(2)_{A'}\times SU(2)_{B'}$, but 
in a different way than discussed there. For the case at hand we take 
$SU(2)_{A'}$ to rotate the first three scalars $\Phi^1$, $\Phi^2$ and 
$\Phi^3$. $SU(2)_{B'}$ will rotate the remaining three scalars, which do not 
appear in $Z$ and therefore we will not find any constraints associate to it. 
Under this breaking, which is discussed in detail in 
Appendix~\ref{app-R2-susy}, the ${\bf4}$ of $SO(6)$ is decomposed into 
the $({\bf2},{\bf2})$ of the broken group.%
\footnote{The breaking in Section~\ref{sec-R4-susy} is such that 
${\bf4}\to ({\bf2},{\bf1})\oplus({\bf1},{\bf2})$.} 
The index $A$ of $SU(4)$ is replaced
by the pair $\ad a$, with the dotted and undotted indices representing 
$SU(2)_{A'}$ and $SU(2)_{B'}$ respectively. The anti-symmetric $\rho^{ij}$ with 
$i,j=1,2,3$ are the generators of
$SU(2)_{A'}$ and can be written in terms of Pauli matrices. In addition we
consider the chiral decomposition of the spinors under
$SU(2)_L\times SU(2)_R$ with indices $\alpha$ and $\alphad$
respectively.

Under this decomposition the chiral and anti-chiral parts of $\epsilon_0$ 
have the indices $\epsilon_{0\,\ad a}^{+\,\alpha}$ and 
$\epsilon_0^{-\,\alphad\ad a}$ and of $\epsilon_1$ they are 
$\epsilon_{1\,\alpha}^{+\,\ad a}$ and $\epsilon^-_{1\,\alphad\ad a}$. 
The most general supersymmetry transformation is then generated by
\begin{equation}
\epsilon_{0\,\ad a}^{+\,\alpha}Q_\alpha^{\ \ad a}
+\epsilon_0^{-\,\alphad\ad a}\bar Q_{\alphad\ad a}
+\epsilon_{1\,\alpha}^{+\,\ad a}S^\alpha_{\ \ad a}
-\epsilon^-_{1\,\alphad\ad a}\bar S^{\alphad\ad a}\,.
\label{general-susy-R2}
\end{equation}

The specific choice of gamma matrices in (\ref{gammachoice}) is
such that
\begin{equation}
\gamma^+_{\alpha\alphad}=i\delta_\alpha^1\delta_{\alphad}^{\dot2}\,,\qquad
\gamma^-_{\alpha\alphad}=i\delta_\alpha^2\delta_{\alphad}^{\dot1}\,,\qquad
\gamma^{+\alphad\alpha}=-i\delta^{\alphad}_{\dot1}\delta^\alpha_2\,,\qquad
\gamma^{-\alphad\alpha}=-i\delta^{\alphad}_{\dot2}\delta^\alpha_1\,.
\end{equation}
Likewise \eqn{rhoij-choice}
\begin{equation}
(\rho^{3+})^{\ad}{}_{\bd}
=\delta^{\ad}_{\dot1}\delta_{\bd}^{\dot2}\,,\qquad
(\rho^{3-})^{\ad}{}_{\bd}
=-\delta^{\ad}_{\dot2}\delta_{\bd}^{\dot1}\,.
\end{equation}
The equation $\gamma^-\epsilon_1=0$ means that for the chiral
component, $\epsilon_{1\,\alpha}^{+\,\ad a}$, the subscript $\alpha$ has 
to be 2 and for the anti-chiral part $\alphad=\dot1$.
The equation $\rho^{3+}\epsilon_1=0$ means that
the superscript $\ad=\dot1$, and as a subscript $\ad=\dot2$.
Therefore
\begin{equation}
\epsilon_{1\,\alpha}^{+\,\ad a}=
\delta_\alpha^2\delta^{\ad}_{\dot1}\epsilon^{+\,a}\,,\qquad
\epsilon^-_{1\,\alphad\ad a}=
\delta_{\alphad}^{\dot1}\delta_{\ad}^{\dot2}\epsilon^-_a\,,
\end{equation}
with arbitrary $\epsilon^{+\,a}$ and $\epsilon^-_a$. Now we can
use the first equation in (\ref{Zplane-susy}) to solve for $\epsilon_0$
\begin{equation}
\epsilon_0^{-\,\alphad\ad a}=
i\delta^{\alphad}_{\dot1}\delta^{\ad}_{\dot2}
\epsilon^{+\,a}\,,\qquad
\epsilon_{0\,\ad a}^{+\,\alpha}=
i\delta^{\alpha}_2\delta_{\ad}^{\dot1}\epsilon^-_a\,,
\end{equation}
Using \eqn{general-susy-R2}, this gives the four independent supersymmetry 
generators
\begin{equation}
\cQ^+_a=\bar Q_{\dot1\dot2a}-iS^2_{\ \dot 1a}\,,\qquad
\cQ^{-\,a}=Q_2^{\ \dot1a}+i\bar S^{\dot1\dot2a}\,.
\end{equation}
Note that the supercharges mix $S$ and $Q$ generators of different
chirality.

The supercharges should commute to symmetries of the operators, which
are the rotation in the transverse plane and $SU(2)_{B'}$ rotations
\begin{equation}
\big\{\cQ^+_a\,,\cQ^{-\,b}\big\}
=-i\delta_a^b(J^2_{\ 2}-\bar{J}^{\dot 1}_{\ \dot 1})-iT^b_{\ a}.
\end{equation}
Indeed the trace part is the rotation in the $(x_3,x_4)$ plane and the
triplet of $a$ and $b$ are the $SU(2)_{B'}$ generators.

\subsection{Perturbative calculation}
\label{sec-pert-R2}

We want to calculate $n$-point correlators of operators built out of the field 
$Z$
\begin{equation}
\Bigl \langle \Tr Z^{J_1}(w_1,\bar{w}_1)\ \Tr Z^{J_2}(w_2,\bar{w}_2)\ \cdots\
\Tr Z^{J_n}(w_n,\bar{w}_n)\, \Bigr \rangle \, ,
\end{equation}
with all $n$-points $(w_i,\bar w_i)$ lying in the plane.

At tree level we should consider all possible contractions of $Z$ 
fields. Now using \eqn{uI-R2}
we have $u_I(\bar w_i)\cdot u_I(\bar w_j)=2(\bar w_i-\bar w_j)^2$. The free-field contractions 
are therefore given by
\begin{equation}
[12]\equiv\vev{Z(w_1,\bar{w}_1)\,Z(w_2,\bar{w}_2)}=
\frac{1}{2\pi^2}\frac{\bar{w}_{12}}{w_{12}}\,,\qquad
w_{ij}\equiv w_i-w_j\,,
\label{hol[12]}
\end{equation}
where as before we suppressed gauge indices. 

This two-point function is equivalent to that of a (matrix valued) field in a 
two-dimensional conformal field theory with conformal weights 
$(\frac{1}{2},-\frac{1}{2})$. We discuss the transformation properties of 
the field $Z$ under twisted conformal symmetries in the next subsection.

The operators $\Tr Z^J$ are chiral primary operators of $\cN=4$ SYM, so the
two and three-point functions do not receive quantum corrections and
are given by considering all possible free-field contractions \eqn{hol[12]}.

Unlike the case of the operators in Section~\ref{sec-R4}, for the operators made of 
the field $Z$ on $\bR^2$, we do not have a general proof for the vanishing 
of the quantum corrections. It may be possible to show that the action is exact 
under the supersymmetries that annihilate $Z$, which would prove this statement. 

Instead we proceed here to study the correlation functions of these operators 
in special cases. First we consider the four-point functions, based on the general 
results of \cite{Arutyunov:2002fh,Arutyunov:2003ae}. Then we turn to some 
specific examples of five and six-point functions of operators of low dimension 
and show by explicit calculations performed in our companion paper
\cite{us} that the {\em one}-loop correction vanishes.

The first interesting quantity is the four-point function of these operators.
As discussed in Section~\ref{PertI}, the key ingredients that appear in this 
calculation are the pairwise contractions \eqn{XYZ}
\begin{equation}
\cX=[12][34]\,,\qquad
\cY=[13][24]\,,\qquad
\cZ=[14][23]\,.
\end{equation}
The two other ingredients are the conformal invariant cross ratios \eqn{cross-ratios-R4}, 
which may also be expressed in terms of a complex number $\mu$
\begin{equation}
s=\frac{x_{12}^2\, x_{34}^2}{x_{13}^2\, x_{24}^2}=\mu\bar\mu\, , \qquad
t=\frac{x_{14}^2\, x_{23}^2}{x_{13}^2\, x_{24}^2}=(1-\mu)(1-\bar\mu)\,.
\label{stmu}
\end{equation}
Using these, the universal polynomial prefactor \eqn{cRI} of 
\cite{Arutyunov:2002fh,Arutyunov:2003ae} takes the factorized form
\begin{align}
\cR&= s\, (\cY-\cX)\, (\cZ-\cX) + t\, (\cZ-\cX)\, (\cZ-\cY) +
(\cY-\cX)\, (\cY-\cZ) \nonumber\\
&=\Big(\mu\,(\cX-\cZ)+\cZ-\cY\Big)\, 
\Big(\bar\mu\,(\cX-\cZ)+\cZ-\cY\Big)\,.
\label{4-pt-mu}
\end{align}
So far this expression does not assume our specific operators, it only
uses the complex representation of the cross-ratios (\ref{stmu}).

In our case $\mu$ can be written explicitly as the cross ratio of the four 
points $w_i$ on the complex plane%
\footnote{In general, any four points sit on a sphere or a plane in $\bR^4$ and
defining $\mu$ by solving (\ref{stmu}) will give the complex conformal
cross-ratio of these points with the natural complex-structure on that
sphere/plane.}
\begin{equation}
\mu=\frac{w_{12}\, w_{34}}{w_{13}\, w_{24}}
\end{equation}

We note now that for our special operators $Z$, the pair-wise contractions 
$\cX$, $\cY$ and $\cZ$ are related to the cross-ratios by
\be
\frac{\cX}{\cY}=\frac{\bar\mu}{\mu}\,,\qquad
\frac{\cZ}{\cY}=\frac{1-\bar\mu}{1-\mu}\,.
\label{pairwise-mu}
\ee
With this we find the `magical' identity
\begin{equation}
\mu\,(\cX-\cZ)+\cZ-\cY=0\,,
\label{Ris0}
\end{equation}
so 
\be
\cR=0\, ,
\ee
and therefore all the four-point functions do not receive any
quantum corrections in perturbation theory!

For correlation functions beyond the four-point function we do not have 
general results. We did calculate, though, several five and six-point functions 
at one-loop order and found that the quantum corrections vanish, suggesting that 
this might be a general property of all $n$-point functions. 

The calculation is done by using the results of \cite{us}, where the one loop 
correction to the $n$-point function of chiral primary operators is written as a 
sum of insertions of an effective four-scalar vertex $D_{ijkl}$ into tree-level 
disc amplitudes \eqn{D-graph}.

Using the complex cross-ratio $\mu$, the function $D_{1234}$ can be written 
as
\be
D_{1234}=\frac{\lambda}{32\pi^2}\,\Phi(s,t)
\big(2\,[13][24]-(2-\mu-\bar\mu)[14][23]-(\mu+\bar\mu)[12][34]\big).
\ee
In our case we can furthermore use \eqn{pairwise-mu} to simplify this to
\be
\label{D-graph-mu}
D_{1234}=-\frac{\lambda}{32\pi^2}\,\Phi(s,t)\,\cY\,
\frac{(\mu-\bar\mu)^2}{\mu(1-\mu)}\,.
\ee

$\Phi(s,t)$ is a transcendental function of the cross-ratios \eqn{Phi}, and therefore 
the sum over different $D_{ijkl}$ insertions in \eqn{4-pt} is over different transcendental 
functions among which there cannot be cancelations. The exception are terms 
with the same four vertices but with a different ordering. It is always true that 
$D_{1234}+D_{1324}+D_{1243}=0$, but using the expression in 
\eqn{D-graph-mu} valid for our operators we find furthermore that $D$ 
satisfies the modular relations
\be
D_{1234}=-\frac{1}{\mu}\,D_{1324}=-\frac{1}{1-\mu}\,D_{1243}\,,
\label{modular}
\ee

Let us now examine the particular example of the minimal five-point function, 
that of operators of dimension two, the general insertion formula \eqn{4-pt} 
gives \cite{us}
\begin{align}
\label{5pt}
\vev{\cO^{u_1}_2\cO^{u_2}_2\cO^{u_3}_2\cO^{u_4}_2\cO^{u_5}_2}_\text{1-loop}
&=-32\,\big ( D_{1234}\big([13][52][45]+[15][53][24]\big)
\\&\hskip-3cm
+D_{1324}\big([12][35][54]+[15][52][34\big)
+D_{1243}\big([14][25][53]+[15][54][23]\big)
\nn\\&\hskip-3cm
+\text{cyclic permutations of $(12345)$}\Big).
\nn
\end{align}
Using the modular property \eqn{modular} allows us to simplify the 
three terms we have written explicitly in \eqn{5pt}, which add up to
\be
\begin{aligned}
D_{1234}\Big(&
[13][52][45]+[15][53][24]
-\mu\big([12][35][54]+[15][52][34\big)
\\&
-(1-\mu)\big([14][25][53]+[15][54][23]\big)\Big)
\end{aligned}
\label{5-pt-DDD}
\ee
By an explicit calculation, plugging in the value of the tree-level contractions 
\eqn{hol[12]}, we find that this sum vanishes. 
Hence there are no one-loop correction to this five-point function.

Furthermore, in \cite{us} several other examples of five-point 
functions of operators of total dimension up to sixteen were calculated 
and it was shown that they 
can always be written as a sum of six terms. One is proportional 
to \eqn{5pt} and the rest are proportional to $\cR$ \eqn{4-pt-mu} (with the 
five different choices of four points). Since these constituents vanish 
for operators made solely of $Z$, the one-loop corrections to all these 
five-point functions vanish. If such a decomposition 
of the five-point amplitude generalizes also for chiral primary operators 
of higher dimension, it would then immediately imply the vanishing 
of the one-loop correction to any five-point function made of $Z$.

We also computed in \cite{us} 
one six-point function, that of six chiral primary operators 
of dimension two. It is written again as a sum similar to \eqn{5pt} and 
by plugging in our choice of operators, using the modular relation 
\eqn{modular} we get a sum of fifteen terms similar to \eqn{5-pt-DDD} 
(but with eighteen terms instead of six and each made of four tree-level 
contractions, instead of three). By direct calculation we found that this 
vanishes. It would be interesting to understand higher-point functions, 
both as to their factorization into $n$-point function of operators of 
dimension two, and to the vanishing of the analogs of \eqn{5-pt-DDD}. 
We leave this for future explorations.

We would like to stress again that unlike the field $C$ of 
Section~\ref{sec-R4}, the correlators 
of operators made of $Z$ are not constant, rather they involve the ratio of 
the anti-holomorphic and holomorphic distances between the points.

\subsection{Twisted symmetry}
\label{sec-R2-twist}

It is clear that in addition to the four supersymmetries calculated in 
Section~\ref{sec-R2-susy}, the field $Z$ is invariant under $J_{34}$, 
the rotation that leaves the plane invariant, as well as under the action of the three
generators of the $R$-symmetry group that act on the three remaining 
scalars $\Phi^4$, $\Phi^5$ and $\Phi^6$, which we dubbed $SU(2)_{B'}$.

Beyond that, being restricted to the plane, $Z$ transforms in
representations of $SL(2,\bC)$, of rigid conformal transformations
on the plane generated by $P_1$, $P_2$, $K_1$, $K_2$, $J_{12}$ and
$D$. Likewise, since it involve $\Phi^1$, $\Phi^2$ and $\Phi^3$,
$Z$ can be classified in terms of the $SU(2)_{A'}$ group
that rotates them, generated by $R_{12}$, $R_{23}$ and $R_{31}$.

Consider the three generators of the holomorphic $SL(2,\bR)$
\begin{equation}
\begin{gathered}
L_1=\frac{1}{2}(P_1-iP_2)\,,\qquad
L_0=\frac{1}{2}(D-iJ_{12})\,,\qquad
L_{-1}=\frac{1}{2}(K_1+iK_2)\,.
\end{gathered}
\end{equation}
These operators act on $Z$ by
\begin{equation}
\begin{gathered}
L_1\,Z=\partial_wZ\,,\qquad
L_0\,Z=w\,\partial_wZ+\frac{1}{2}Z\,,\qquad
L_{-1}\,Z=w^2\,\partial_wZ+wZ\,.
\end{gathered}
\end{equation}
$Z$ therefore transforms as a weight $1/2$ primary field of this group.

Since some of the other symmetry generators do not close on $Z$, it will 
prove useful to define two more fields made of the same three scalars
\begin{equation}
Y=-i\bar{w}\Phi^1+\bar{w}\Phi^2-i\Phi^3\,,\qquad
W=-i\Phi^1+\Phi^2\,.
\label{YW}
\end{equation}
The transformation rules of $Y$ and $W$  under $L_i$ are identical 
to that of $Z$. 
This is clearly the same behavior as for any of the scalar fields,
since there is no explicit $w$ dependence in
the definitions of $Z$, $Y$ and $W$

The rest of the symmetry generators can be organized as
\begin{align}
R_+&=-i(R_{23}+iR_{31})\,,\qquad&
R_0&=iR_{12}\,,\qquad&
R_-&=i(R_{23}-iR_{31})\,,
\\
\bar{L}_1&=\frac{1}{2}(P_1+iP_2)\,,\qquad&
\bar{L}_0&=\frac{1}{2}(D+iJ_{12})\,,\qquad&
\bar{L}_{-1}&=\frac{1}{2}(K_1-iK_2)\,,
\end{align}
Their action on the fields $Z$, $Y$ and $W$ are not too simple 
and are given in the appendix, see \eqn{R2-R-action} \eqn{R2-barL-action}.

A natural thing to try is to take the linear combination of $\bar L$ 
and $R$. Consider for example
\begin{equation}
\dot{L}_1=\bar{L}_1+R_+\,,\qquad
\dot{L}_0=\bar{L}_0+R_0\,,\qquad
\dot{L}_{-1}=\bar{L}_{-1}+R_-\,.
\end{equation}
Their action on $Z$ is given by
\be
\dot{L}_1\,Z=\partial_{\bar{w}}Z\,,\qquad
\dot{L}_0\,Z=\bar{w}\,\partial_{\bar{w}}Z-\frac{1}{2}Z\,,\qquad
\dot{L}_{-1}\,Z=\bar{w}^2\,\partial_{\bar{w}}Z-\bar{w}Z\,.
\ee
$Z$ therefore transforms as a weight $-1/2$ field of this
twisted anti-holomorphic $SL(2,\bR)$.
The action on $Y$ and $W$ is given in \eqn{R2-dotL-action}. 
$Y$ has weight $1/2$ and $W$ has weight $3/2$, but they are
not primaries, since there are additional terms in the action
of $\dot{L}_{-1}$.

It will turn out that a different combination of the anti-holomorphic symmetry
generators and rotations is related to supersymmetries preserved by the 
operators $Z$. These are
\begin{equation}
\hat{L}_1=\bar{L}_1+\frac{1}{2}R_+\,,\qquad
\hat{L}_0=\bar{L}_0+\frac{1}{2}R_0\,,\qquad
\hat{L}_{-1}=\bar{L}_{-1}+\frac{1}{2}R_-\,.
\end{equation}
Note that because of the factor of $1/2$ those generators do not close 
onto themselves, and do not form an $SL(2,\bR)$ algebra.

The action of these operators on $Z$ is
\be
\hat{L}_1\,Z=\partial_{\bar{w}}Z-Y\,,\qquad
\hat{L}_0\,Z=\bar{w}\,\partial_{\bar{w}}Z-\bar wY\,,\qquad
\hat{L}_{-1}\,Z=\bar{w}^2\,\partial_{\bar{w}}Z-\bar w^2Y\,.
\label{L-hat-action}
\ee
Under this twisting $Z$ has dimension zero, but has these extra terms 
proportional to $Y$ in the action of $\hat L$. 
The actions on $Y$ and $W$ are given in \eqn{app-L-hat-action}, where 
$Y$ has dimension $1/2$ and $W$ dimension one.

To see how these symmetry generators come about, 
consider the anti-commutators of $\cQ^\pm$ with all the
other supercharges which will generate some of the bosonic symmetries
of the theory. Most of these symmetries will map our operators to
others, taking them away from the $(x_1,x_2)$ plane or turning on
the three remaining scalars. But the following combinations map
our operators to themselves
\begin{align}
\big\{\cQ^+_a\,,iQ_2^{\ \dot1a}+\bar S^{\dot1\dot2a}\big\}
&=2\big(J^2_{\ 2}+\bar{J}^{\dot 1}_{\ \dot 1}+D\big)
+\dot T^{\dot1}_{\ \dot1}-\dot T^{\dot2}_{\ \dot2}
=2(D+iJ_{12}+iR_{12})
=4\hat L_0\,,
\nn\\
\big\{\cQ^+_a\,,-iQ_2^{\ \dot2a}\big\}
&=-2iP_{2\dot1}-\dot T^{\dot2}_{\ \dot1}
=P_1+iP_2-i(R_{23}+iR_{31})
=2\hat L_1\,,
\nn\\
\big\{\cQ^+_a\,,-\bar S^{\dot1\dot1a}\big\}
&=2iK^{\dot12}+\dot T^{\dot1}_{\ \dot2}
=K_1-iK_2+i(R_{23}-iR_{31})
=2\hat L_{-1}\,.
\label{exact-symmetries-R2}
\end{align}
Similar expressions exist for $\cQ^{-a}$ giving the same combinations
of symmetry generators on the right-hand side. Note that these symmetries
include both space-time generators and $R$-rotations of $SU(2)_{A'}$
and are the second twisting discussed above. Their action on the
fields $Z$, $Y$ and $W$ are given in \eqn{L-hat-action} and 
\eqn{app-L-hat-action}.

These twisted symmetry generators can be used to find extra 
relations among the $n$-point function of operators with $\Tr Z^J$ which 
are valid in the quantum theory.

It is instructive to consider the contractions of $Z$ as well as $Y$ and $W$ \eqn{YW}
(again suppressing the gauge group indices)
\begin{align}
\vev{Z(w_1,\bar{w}_1)\,Z(w_2,\bar{w}_2)}&=
\frac{1}{2\pi^2}\frac{\bar{w}_{12}}{w_{12}}\,,\quad&
\vev{Y(w_1,\bar{w}_1)\,Y(w_2,\bar{w}_2)}&=
-\frac{1}{4\pi^2}\frac{1}{w_{12}\,\bar{w}_{12}}\,,
\nn\\
\vev{Y(w_1,\bar{w}_1)\,Z(w_2,\bar{w}_2)}&=\frac{1}{2\pi^2}\frac{1}{w_{12}}
\,,\quad&
\vev{W(w_1,\bar{w}_1)\,Z(w_2,\bar{w}_2)}
&=\frac{1}{2\pi^2}\frac{1}{w_{12}\,\bar{w}_{12}}\,,
\nn\\
\vev{W(w_1,\bar{w}_1)\,Y(w_2,\bar{w}_2)}&= 0\,,\quad&
\vev{W(w_1,\bar{w}_1)\,W(w_2,\bar{w}_2)}&=0\, .
\label{tree-contractions}
\end{align}

Consider the action of $\cQ^+_a$ on the correlator of any number
of $\Tr Z^J$ operators and one arbitrary local operator $\cO$
\begin{equation}
\cQ^+_a\vev{\cO\ \Tr Z^{J_2}\cdots\Tr Z^{J_n}\cdots}
=\vev{\cQ^+_a\cO\ \Tr Z^{J_2}\cdots\Tr Z^{J_n}\cdots}
\end{equation}
$\cQ^+_a$ commutes with all the $Z$'s and the overall expression
vanishes, by a Ward-Takahashi identity.

Now take $\cO=\frac{1}{2J_1}\,Q_2^{\ \dot2a}\Tr Z^{J_1}$.
Since we saw \eqn{exact-symmetries-R2} that 
$2\hat L_1=\big\{\cQ^+_a\,,-iQ_2^{\ \dot2a}\big\}$
and it commutes with the $Z$'s, we have
\begin{equation}
-i\cQ_a^+\cO=\frac{1}{J_1}\,\hat L_1\Tr Z^{J_1}
=\Tr\left[(\partial_{\bar w}Z-Y) Z^{J_1-1}\right]\,.
\end{equation}
Thus we find the following relation for the four-point 
function with one $Y$ insertion
\begin{equation}
\vev{\Tr[YZ^{J_1-1}]\,\Tr Z^{J_2}\,\Tr Z^{J_3}\,\Tr Z^{J_4}}
=\frac{1}{J_1}\partial_{\bar w_1}
\vev{\Tr Z^{J_1}\,\Tr Z^{J_2}\,\Tr Z^{J_3}\,\Tr Z^{J_4}}\,.
\end{equation}
As we have proven in Section~\ref{sec-pert-R2}, the four-point function 
on the right-hand side is given by the free contractions of the different $Z$'s 
and from this we derived an {\em exact} expression for the correlator 
on the left-hand side as well. Similar statements 
would hold for higher $n$-point functions if indeed these
are not renormalized either.

To illustrate this type of relation in a particularly simple example, 
for the two-point function we know from (\ref{tree-contractions}) that
\begin{equation}
\vev{Z(w_1,\bar{w}_1)\,Z(w_2,\bar{w}_2)}=
\frac{1}{2\pi^2}\frac{\bar{w}_{12}}{w_{12}}\,,\qquad
\vev{Y(w_1,\bar{w}_1)\,Z(w_2,\bar{w}_2)} =\frac{1}{2\pi^2}\frac{1}{w_{12}}\,,
\end{equation}
and indeed
$\vev{Y(w_1,\bar w_1)\,Z(w_2,\bar w_2)}
=\partial_{\bar w_1}\vev{Z(w_1,\bar w_1)\,Z(w_2,\bar w_2)}$\,.

The twisted symmetry generators $\hat L_i$ can be used to derive more 
such relations between correlation functions.

\section{Discussion}
\label{sec-discussion}

In this paper we introduced the notion of ``superprotected $n$-point 
function'', the correlation function of operators all sharing 
supersymmetries. We focused on two main examples: In 
Section~\ref{sec-R4} operators constructed of all six scalars and at 
general position in $\bR^4$, and in Section~\ref{sec-R2} operators 
constructed out of three real scalars and restricted to a plane.

The operators have explicit spatial dependence and in the example 
of Section~\ref{sec-R4} this renders their tree-level correlation functions 
space-independent. Thus these correlation functions are given by a 
zero-dimen\-sional Gaussian matrix model. Furthermore we provided 
different evidence for the absence of perturbative corrections to 
these observables. The most elegant argument is that given by 
de Medeiros et al. \cite{deMedeiros:2001kx}, who showed that the 
$\cN=4$ action is exact under the supersymmetries that annihilate 
these operators, up to instanton terms. Beyond this somewhat formal 
argument we checked this cancellation using an explicit expression 
for the one-loop correction to all $n$-point functions of chiral 
primary operators, published in an accompanying paper \cite{us}. 
In addition we relied on the general structure of the four-point function 
of chiral primary operators \cite{Arutyunov:2002fh,Arutyunov:2003ae} 
which implies the all-loop cancelation of quantum corrections, even 
including instantons \cite{Arutyunov:2000im}.

The operators in Section~\ref{sec-R2} have a different spatial 
dependance and consequently more complicated correlation functions. 
The free contractions are those of a 
two-dimensional CFT with matrix fields of dimension 
$(\frac{1}{2},-\frac{1}{2})$. Again, we checked the quantum corrections 
in a variety of ways, the all-loop corrections to all four point functions 
and the explicit one-loop correction to some five-point functions and 
one six-point function. In all these cases the quantum corrections 
vanished, leading one to believe that again these $n$-point functions 
are given by this free theory.

There exists another class of $n$-point functions that do not receive 
quantum corrections, the extremal correlators \cite{D'Hoker:1999ea}. 
These correlation functions are such that the weights of the operators 
allow only very simple Feynman diagrams to contribute and exclude 
quantum corrections. Our constructions are based on a very different 
principle; the weights are completely arbitrary, but the type of operator 
is correlated with its space-time position. The simplicity is a consequence 
of the supersymmetry shared by all the operators.

It would clearly be desirable to have rigorous proofs  that none of the 
correlation functions studied in this paper receive perturbative 
corrections. For the case discussed in Section~\ref{sec-R4}, this is 
done in \cite{deMedeiros:2001kx} by showing that the action is $\cQ$ 
exact, up to instanton terms. It would be interesting to try to show the 
same for the operators in Section~\ref{sec-R2}. Furthermore, since 
the proof applies only to the perturbative series, it suggests that there 
could be instanton corrections to the $n$-point functions and perhaps 
they are also computable.

Another question we have not touched on is regarding the string duals 
of these $n$-point functions. Four-point functions have been calculated 
in $AdS_5\times S^5$ \cite{D'Hoker:1999pj,D'Hoker:1999ni,Arutyunov:2000py} 
and it is known that the result is also 
proportional to the universal polynomial prefactor $\cR(s,t;\cX,\cY,\cZ)$ 
\eqn{cRI} of \cite{Arutyunov:2002fh,Arutyunov:2003ae}. Hence the 
quantum corrections in string theory also cancel for the four-point 
functions. It would still be nice to have explicit calculations for our 
examples, since they most likely are much simpler than a generic 
four-point function calculation (which is quite complicated). Is there 
some way to organize the calculation which brings out the fact that 
the full result localizes to free graphs? Furthermore, would it be 
possible to calculate in $AdS$ higher-point functions for these 
operators?

Going beyond the specific examples studied in this paper, one could 
ask the same question regarding any $n$-point function where all 
the operators share some supersymmetries. Are all such correlation 
functions protected? There are many examples where 
BPS Wilson loop operators are \cite{Erickson:2000af, 
Drukker:2000rr, Drukker:2005kx,Yamaguchi:2006tq,  Gomis:2006sb, 
Hartnoll:2006is, 1/4-giant, Gomis:2006im, Pestun:2007rz}. Likewise 
the known examples of correlation functions of local operators and 
Wilson loops sharing some supersymmetries are given by summing 
free propagators \cite{Semenoff:2001xp,Zarembo:2002ph, 
Pestun:2002mr, Okuyama:2006jc, Giombi:2006de, Semenoff:2006am, 
Gomis:2008qa}. One could also find Wilson loops that share 
supersymmetry with the $n$-point functions discussed in this 
paper \cite{Pestun}. As another example, the slightly more exotic surface 
operators \cite{Gukov:2006jk} seem to have very simple correlation 
functions with Wilson loops and local operators when they all share 
supersymmetry \cite{Drukker:2008wr,Koh:2008kt}.

In fact, another family of local operators that share supersymmetry 
can be derived from taking infinitesimal Wilson loops. There are 
two known examples of families of Wilson loop operators which all 
share some supercharges \cite{Zarembo:2002an,Drukker:2007dw}. 
While those papers concentrate on the expectation value of a single 
Wilson loop, there is no impediment to take more than one---that 
configuration is still supersymmetric. When shrinking all the Wilson 
loops to small size, one ends up with local operators which can 
serve as another realization of the ideas put forth in this paper. 
The Wilson loops are made by 
including special scalar couplings in addition to the gauge connection. 
The resulting local operators will include the field strength, derivatives 
and commutators of scalar fields, which can all be represented in 
terms of some modified covariant derivative like
\be
\begin{aligned}
\widehat\cD_\mu&=\partial_\mu-iA_\mu+\Phi_\mu\,,
\\
\widetilde\cD_\mu&=\partial_\mu-iA_\mu+\Phi^+_{\mu\nu}x^\nu\,.
\end{aligned}
\ee
These correspond to the two examples, where the scalar fields get 
assigned space-time indices of a vector and self-dual tensor in a 
natural way \cite{Zarembo:2002an,Drukker:2007dw}.

In the case of the loops constructed by Zarembo \cite{Zarembo:2002an}, 
the expectation values are always unity 
\cite{Guralnik:2003di,Guralnik:2004yc,Dymarsky:2006ve}, and it is 
reasonable to expect that this would be true also in the limit. The 
second example, that in\cite{Drukker:2007dw} is more complicated 
and one would expect the $n$-point function of the infinitesimal 
Wilson loops to be non-zero. In particular, when the loop is restricted 
to an $S^2$ in space-time there is some evidence showing that they 
are equal to a perturbative calculation in two-dimensional Yang-Mills theory 
\cite{Drukker:2007yx,Drukker:2007qr} (see also 
\cite{Bassetto:2008yf,Young:2008ed}). Are the correlation functions of 
the infinitesimal loops then given by the correlators of single plaquette 
operators in two-dimensional Yang-Mils?

These examples, including the ones studied in this paper are surely 
not the only ones. For local operators, as mentioned before, any three 
chiral primary operators will share some supercharges. It is reasonable 
to expect that on the line (or circle) spanned by these operators one 
could place more local operators that share the same supercharges 
as the original three. Will all such objects have vanishing quantum 
corrections? It is possible that the tools we used in Section~\ref{sec-R2} 
would apply also there. In checking for the cancelation of the one-loop 
corrections an important property was the way the interaction vertex 
depended on the complex cross-ratio \eqn{modular}. For four operators 
on a line there is only one real cross-ratio, so it is possible that 
similar relations will also hold.

Beyond a single line, one can ask whether there are other examples 
of families of operators on submanifolds of space-time, like $\bR^2$ 
in our second example, that share some supercharges, and whether 
they receive quantum corrections. One useful tool may be the universal 
polynomial function $\cR$ \eqn{cRI}. In both of our examples it vanished, 
proving that the four point functions do not get renormalized. One can 
therefore ask for which collection of points, or submanifold of space-time 
and for which operators does $\cR$ vanish. In these cases will the 
operators necessarily share some supercharges? Under what conditions 
would it be possible to add operators to make $n$-point functions with 
vanishing quantum corrections, and is there a generalization of $\cR$ 
to these cases (see also \cite{us}).

As we touched on in the text, the supersymmetry shared by the families 
of operators we constructed lead to some bosonic ``twisted'' symmetries 
that relate different correlation functions to each-other. It would be 
interesting to understand the scope of these symmetries and find all 
possible correlation functions of other operators, involving fermions, 
derivatives and gauge fields and which are related to the ones we 
have calculated---and therefore are also ``superprotected''.

Our results advocate the point of view where one should not necessarily 
regard local operators as the basic objects and $n$-point functions merely 
as their correlators. The $n$-point functions may have more of an 
independent meaning. One example of this dual point of view are 
classical geodesics in $AdS$ space---they calculate the two-point 
functions of dual operators. In particular, in all the examples that we 
studied we investigated the amount of supersymmetry preserved by 
all the objects in the correlation function, not each separately.

A very interesting spin-off would be to try to build upon our ``superprotected'' 
three-point functions to understand the interaction of non-BPS operators. 
In the same way that the spectrum of local operators is understood in terms 
of magnon excitations over a supersymmetric ground state, one could 
put magnons on top of three long operators which share supersymmetry 
and study their interactions. We find the operators in Section~\ref{sec-R4} 
particularly promising candidates for the ground state, since their 
correlation functions have trivial spatial dependence.

\subsection*{Acknowledgments}
We would like to thank
Ofer Aharony, Niklas Beisert, Jaume Gomis, Volker Schomerus, Stefan Theisen and 
Donovan Young for stimulating discussions. We thank the Galileo-Galilei Institute in Firenze 
for hospitality during the course of parts of this work. N.D. also acknowledges DESY, Hamburg, 
and SISSA, Trieste and NBI, Copenhagen 
for hospitality and the INFN for partial financial support.  
This work was supported by the Volkswagen-Foundation.

\appendix
\section{Notations and the superalgebra}
\label{app-notation}

This appendix summarizes our conventions
for the $\cN=4$ superconformal algebra $PSU(2,2|4)$ following
\cite{Beisert:2004ry}. 
The two ways of breaking the $R$-symmetry group 
$SU(4)\to SU(2)\times SU(2)$ are then explained in the 
following appendices.

We denote by $J^{\alpha}_{\ \beta}$, $\bar
J^{\alphad}_{\ \betad}$ the generators of the $SU(2)_L
\times SU(2)_R$ Lorentz group, and by  $R^A_{\ B}$ the 15
generators of the $R$-symmetry group $SU(4)$. The remaining bosonic
generators are the translations $P_{\alpha\alphad}$, the
special conformal transformations $K^{\alpha\alphad}$ and the
dilatation $D$. Finally the 32 fermionic generators are the
Poincar\'e supersymmetries $Q^A_{\alpha}$, $\bar Q_{\alphad A}$ and
the superconformal supersymmetries $S^{\alpha}_A$,
$\bar S^{\alphad A}$.

The commutators of any generator $\cG$ with
$J^{\alpha}_{\ \beta}$, $\bar J^{\alphad}_{\ \betad}$
and $R^A_{\ B}$ are canonically dictated by the index structure
\begin{equation}
\begin{gathered}{}
\big[J^{\alpha}_{\ \beta}\,,\cG_\gamma\big]
=\delta^\alpha_\gamma \cG_\beta
-\frac{1}{2}\delta^\alpha_\beta \cG_\gamma\,,\qquad
\big[J^{\alpha}_{\ \beta}\,,\cG^\gamma\big]
=-\delta_\beta^\gamma \cG^\alpha
+\frac{1}{2}\delta^\alpha_\beta \cG^\gamma\,,
\\
\big[\bar{J}^{\alphad}_{\ \betad}\,,\cG_{\gammad}\big]
=\delta^{\alphad}_{\gammad} \cG_{\betad}
-\frac{1}{2}\delta^{\alphad}_{\betad} \cG_{\gammad}
\,,\qquad
\big[\bar{J}^{\alphad}_{\ \betad}\,,\cG^{\gammad}\big]
=-\delta_{\betad}^{\gammad} \cG^{\alphad}
+\frac{1}{2}\delta^{\alphad}_{\betad} \cG^{\gammad}
\\
\big[R^A_{\ B}\,,\cG_C\big]=\delta^A_C \cG_B
-\frac{1}{4}\delta^A_B \cG_C\,,\qquad
\big[R^A_{\ B}\,,\cG^C\big]=-\delta_B^C \cG^A
+\frac{1}{4}\delta^A_B \cG^C\,.
\end{gathered}
\end{equation}
while commutators with the
dilatation operator $D$ are given by $\big[D\,,\cal G \big] =
\text{dim}(\cal G)\,\cal G$, where $\text{dim}(\cal G)$ is the
dimension of the generator $\cal G$.

The remaining non-trivial commutators are
\begin{equation}
\begin{aligned}
&\big\{ Q^A_{\alpha}\,, \bar Q_{\alphad B} \big\}=
\delta^A_B P_{\alpha\alphad}\,, \qquad \quad \big\{
S^{\alpha}_A \,,\bar S^{\alphad B} \big\}
= \delta^B_A K^{\alpha\alphad}\,, \\
&\big[K^{\alpha\alphad}\,, Q^A_{\beta} \big]=
\delta^{\alpha}_{\beta}\bar S^{\alphad A}\,, \qquad \quad ~
\big[K^{\alpha\alphad}\,, \bar Q_{\betad A} \big] =
\delta^{\alphad}_{\betad} S^{\alpha}_A\,, \\
&\big[P_{\alpha\alphad}\,, S^{\beta}_{A} \big]= -
\delta^{\beta}_{\alpha}\bar Q_{\alphad A}\,, \qquad \quad
\!\!\!~ \big[P_{\alpha\alphad}\,, \bar S^{\betad A}
\big] =
-\delta^{\betad}_{\alphad} Q_{\alpha}^A\,, \\
&\big\{ Q^A_{\alpha} \,, S^{\beta}_{B} \big\} = \delta^A_B
J^{\beta}_{\,\,\alpha} +
\delta^{\beta}_{\alpha} R^A_{\ B} + \frac{1}{2} \delta^A_B \delta^{\beta}_{\alpha} D\,, \\
&\big\{ \bar Q_{\alphad A} \,, \bar S^{\betad B} \big\}
= \delta^B_A \bar{J}^{\betad}_{\,\,\alphad} -
\delta^{\betad}_{\alphad} R^B_{\ A} + \frac{1}{2} \delta^B_A \delta^{\betad}_{\alphad} D\,, \\
&\big[ K^{\alpha\alphad}\,,
P_{\beta\betad}\big]=\delta^{\alphad}_{\betad}
J^{\alpha}_{\ \beta} + \delta^{\alpha}_{\beta}
\bar{J}^{\alphad}_{\ \betad} + \delta^{\alpha}_{\beta}
\delta^{\alphad}_{\betad} D\,.
\end{aligned}
\label{bigalgebra1}
\end{equation}

So far we have written the algebra in spinor notations, but we find it 
useful also to transform to vector notations. To that end we take the 
following choice of Euclidean gamma matrices for $\bR^4$, where 
$\tau^i$ are the usual Pauli matrices
\be
\label{gammachoice}
\gamma^i=
\begin{pmatrix}
0&(\sigma_i)_{\alpha \alphad}\cr
(\bar\sigma^i)^{\alphad \alpha}&0
\end{pmatrix}
=
\begin{pmatrix}
0&i\tau^i\cr
-i\tau^i&0
\end{pmatrix}
\qquad
\gamma^4=
\begin{pmatrix}
0&(\sigma_4)_{\alpha \alphad}\cr
(\bar\sigma^4)^{\alphad \alpha}&0
\end{pmatrix}
=
\begin{pmatrix}
0&\bI\cr
\bI&0
\end{pmatrix}
\ee

$SU(2)$ indices can be raised and lowered by using the appropriate
epsilon tensor, for which we adopt the conventions
\begin{equation}
\label{epsconvention}
\cG^{r}= \varepsilon^{rs}\cG_s\,, \quad
\cG_r= \varepsilon_{rs}\cG^s\,;\qquad
\varepsilon^{rs}=
\begin{pmatrix}
0&1\cr-1&0 
\end{pmatrix}\,,
\quad
\varepsilon_{rs}=
\begin{pmatrix}
0&-1\cr1&0
\end{pmatrix}\,.
\end{equation}
where the indices $r,s$ belong to any $SU(2)$ group.
Indeed 
$(\bar\sigma^\mu)^{\alphad\alpha}=
\epsilon^{\alphad\betad}\, \epsilon^{\alpha\beta}\, (\sigma^\mu)_{\beta\betad}$.

We note the contraction relations
\be
(\sigma^\mu)_{\alpha\alphad}\, (\bar\sigma^\mu)^{\betad\beta}= 2\, \delta^\betad_\alphad\,
\delta^\beta_\alpha\qquad
\Tr(\sigma^\mu\bar\sigma^\nu) = 2\, \delta^{\mu\nu} \, .
\ee
The gamma matrices with anti-symmetric indices are
\be
\sigma^{\mu\nu}=\frac{1}{2}
\left(\sigma^\mu\bar\sigma^\nu-\sigma^\nu\bar\sigma^\mu\right)\,,\qquad
\bar\sigma^{\mu\nu}=\frac{1}{2}
\left(\bar\sigma^\mu\sigma^\nu-\bar\sigma^\nu\sigma^\mu\right)\,.
\ee

We may now define
\begin{equation}
\begin{aligned}
P^\mu&= P_{\alpha\alphad}\, (\bar \sigma^\mu)^{\alphad\alpha}\,,& \qquad
P_{\alphad\alpha}&=\frac{1}{2}\,(\sigma^\mu)_{\alpha\alphad}\,P_\mu\,,\\
K^\mu&= K^{\alphad\alpha}\, (\sigma^\mu)_{\alpha\alphad}\,,& \qquad
K^{\alphad\alpha}&= \frac{1}{2}\,(\bar\sigma_\mu)^{\alphad\alpha}\,
K^\mu\,,\\
J^{\mu\nu}&= \frac{1}{2}
\left(J^\alpha{}_\beta\,\sigma^{\mu\nu})_\alpha{}^\beta
-\bar J^\alphad{}_\betad\, (\bar\sigma^{\mu\nu})^\betad{}_\alphad\, 
\right).&
\end{aligned}
\label{conf-generators}
\end{equation}
Using the commutation relations (\ref{bigalgebra1}) and contracting
the relevant $\sigma^\mu$ and $\bar\sigma^\nu$ we get the commutators
in $SO(4)$ language
\begin{equation}
\begin{aligned}
\big[K^\mu,\,P^\nu\big]&=2\, (J^{\mu\nu} + \delta^{\mu\nu}\, D )\\
\big[J^{\mu\nu},\,P^\rho\big]&=
\delta^{\mu\rho}\,P^\nu-\delta^{\nu\rho}\,P^\mu\,,\\
\big[J^{\mu\nu},\, J^{\rho\sigma}\big]
&=\delta^{\mu\rho}\, J^{\nu\sigma}-\delta^{\nu\rho}\, J^{\mu\sigma}
+\delta^{\mu\sigma} \, J^{\rho\nu}-\delta^{\nu\sigma} \, J^{\rho\mu}\,.
\end{aligned}
\label{commute}
\end{equation}

These commutation relations can be realized by the following definition of the 
action of the symmetry generators on scalar fields
\begin{equation}
\begin{aligned}
P_\mu\,\Phi^i&=\partial_\mu\Phi^i\,,\\
J_{\mu\nu}\,\Phi^i
&=(x_\mu\partial_\nu-x_\nu\partial_\mu)\Phi^i\,,\\
D\,\Phi^i&=(x^\mu\partial_\mu+\Delta)\Phi^i\,,\\
K_\mu\,\Phi^i &=(2x_\mu x^\nu\partial_\nu+2\Delta x_\mu-x^2\partial_\mu)
\Phi^i\,.
\end{aligned}
\label{symmetry}
\end{equation}
Noting that to calculate the commutators the derivatives act on fields, 
and not directly on the coordinates, one gets the commutation relations
\eqn{commute}.

The action of the R-symmetry generators on the scalar fields can be written as
\begin{equation}
R_{ij}\,\Phi^k=\delta_i^k\Phi_j-\delta_j^k\Phi_i\,,
\label{R-symmetry}
\end{equation}
which gives the algebra
\begin{equation}
\big[R_{ij}\,,R_{kl}\big]=\delta_{ik}R_{jl}
-\delta_{jk}R_{li}+\delta_{il}R_{kj}
-\delta_{jl}R_{ik}\,,
\label{R-commutator}
\end{equation}
We choose specific notations for the R-symmetry generators in the 
following two appendices, once we break $SO(6)$ to $SU(2)\times SU(2)$ 
in the two ways appropriate for the different local operators discussed 
in the text.

\section{Symmetry breaking for example I}
\label{app-R4-susy}

The construction of the operators in Section~\ref{sec-R4} involves
an identification of the full $SO(5,1)$ conformal group and the
$R$-symmetry group. The supercharges, which transform in two 
bi-spinor representations of those groups may be decomposed, after 
the identification, to two adjoints and two singlets of the diagonal group. 
The supercharges preserved by the field $C$ are the singlets. 
The standard notations have the $SO(4)$ Euclidean Lorentz group written 
as $SU(2)_L\times SU(2)_R$
with the spinors in the $({\bf2},{\bf1})\oplus({\bf1},{\bf2})$
representations, labeled by the indices $\alpha$ and $\alphad$.
Therefore, to describe the supersymmetry preserved by the operators
on $\bR^4$ it is useful to consider the breaking of the
$SU(4)$ $R$-symmetry group to $SU(2)_A\times SU(2)_B$ such that
the spinor representation becomes
${\bf4}\to({\bf2},{\bf1})\oplus({\bf1},{\bf2})$. We will use indices
$\ad$ for $SU(2)_A$ and $a$ for $SU(2)_B$. Note that
a different breaking is used for the operators on $\bR^2$
and will be described below in Appendix~\ref{app-R2-susy}.

Under this breaking the supergroup generators
\begin{equation}
\left(\begin{array}{cc|c}
J_\alpha^{\ \beta}+\frac{1}{2}\delta^\beta_\alpha D\
&P_{\alpha\betad}
&\ Q^B_\alpha\cr
-K^{\alphad\beta}
&-\bar J^\alphad_{\ \betad}-\frac{1}{2}\delta_\betad^\alphad D\
&\ -\bar S^{\alphad B}\cr
\hline
\vbox{\vskip7mm}
S_A^\beta
&\bar Q_{\betad A}
&R^B_{\ A}
\end{array}\right)
\label{beisert-matrix}
\end{equation}
are decomposed as
\begin{equation}
\left(\begin{array}{cc|cc}
J_\alpha^{\ \beta}+\frac{1}{2}\delta^\beta_\alpha D\
&P_{\alpha\betad}
&Q_\alpha^b
&\dot Q_{\alpha\bd}
\cr
-K^{\alphad\beta}
&-\bar J^\alphad_{\ \betad}-\frac{1}{2}\delta_\betad^\alphad D\
&-\bar S^{\alphad b}
&-\dot{\bar S}^\alphad_\bd
\cr\hline
\vbox{\vskip6mm}
S_a^\beta
&\bar Q_{\betad a}
&\ R^b_{\ a}+\frac{1}{2}\delta^b_a\dot D
&\dot P_{a\bd}
\cr
-\dot S^{\beta\ad}
&-\dot{\bar Q}_\betad^\ad
&-\dot K^{\ad b}
&-\dot R^\ad_{\ \bd}-\frac{1}{2}\delta^\ad_\bd\dot D
\end{array}\right)
\label{broken-matrix}
\end{equation}

This decomposition of the $PSU(2,2|4)$ algebra
into $SU(2)_L\times SU(2)_R\times SU(2)_A\times SU(2)_B$ is realized 
in a very simple way using the osclillator picture of \cite{Beisert:2003jj}.
One starts with two pairs of bosonic oscillators ($\alpha,\alphad=1,2$)
\be
\big[a^\alpha,\,a^\dagger_\beta\big]=\delta^\alpha_\beta\,,\qquad
\big[b^\alphad,\,b^\dagger_\betad\big]=\delta^\alphad_\betad\,,
\ee
and four fermionic oscillators ($A=1,2,3,4$)
\be
\big\{c^A,\,c^\dagger_B\big\}=\delta^A_B\,.
\ee
Then one rewrites the fermionic generators in terms of the two pairs
$c^a$ and $d^\ad$ (with $a,\ad=1,2$ and standard anti-commutators)
\begin{equation}
c^A=(\, c^1,c^2,d^\dagger_{\dot 1} ,d^\dagger_{\dot 2}\, ) \qquad
c^\dagger_A=(\, c^\dagger_{1} ,c^\dagger_{2},d^{\dot 1}, d^{\dot 2}\, )
\end{equation}
The bosonic generators of the algebra are made either of two bosonic 
oscillators (giving the conformal part) or two fermionic ones (giving the 
$R$-symmetry part)
\begin{equation}
\begin{aligned}
J^\alpha{}_\beta & = a^\dagger_\beta\, a^\alpha -
\half \delta^\alpha_\beta\, a^\dagger_\gamma\,a^\gamma \qquad
&{\bar J}^\alphad{}_\betad &= b^\dagger_\betad\, b^\alphad
- \half \delta^\alphad_\betad\, b^\dagger_\gammad\, b^\gammad\\
P_{\alpha\betad}&= a^\dagger_\alpha\, b^\dagger_\betad \qquad\qquad
K^{\alpha\betad}= a^\alpha\, b^\betad \qquad
&D &= 1+\half (a^\dagger_\gamma a^\gamma +b^\dagger_\gammad b^\gammad ) \\
R^a{}_b & = c^\dagger_b\, c^a -\half \delta^a_b\, c^\dagger_c\,c^c \qquad
&\dot R^\ad{}_\bd  &= d^\dagger_\bd\, d^\ad
-\half \delta^\ad_\bd\, d^\dagger_\cd\, d^\cd \\
{\dot P}_{a\bd}&= c^\dagger_a\, d^\dagger_\bd \qquad\qquad
{\dot K}^{a\bd}= c^a\, d^\bd \qquad
&{\dot D}&= -1+\half (c^\dagger_c c^c +d^\dagger_\cd d^\cd )
\end{aligned}
\end{equation}
The fermionic generators of the superalgebra can then be written as
\begin{equation}
\begin{aligned}
Q^a{}_\alpha &= a^\dagger_\alpha\, c^a\,, \qquad&
\bar Q_{a\alphad} &= b^\dagger_\alphad\, c^\dagger_a\,, \qquad&
S^\alpha{}_a &= c^\dagger_a\, a^\alpha\,, \qquad&
\bar S^{\alphad a}&= b^\alphad\, c^a\,, \\
\dot Q_{\ad\alpha}&= a^\dagger_\alpha\,d^\dagger_\ad\,,  \qquad&
\dot{\bar Q}^\ad{}_\alphad &= -b^\dagger_\alphad\,d^\ad\,,\qquad&
\dot S^{\alpha \ad}& = -a^\alpha\, d^\ad\,,\qquad&
\dot {\bar S}^\alphad{}_\ad &= d^\dagger_\ad\,b^\alphad\,.
\end{aligned}
\end{equation}

Some of their commutators are
\begin{equation}
\begin{aligned}
&\big\{ Q^a_{\alpha},\, \bar Q_{b\alphad} \big\}
=\delta^a_b P_{\alpha\alphad},\, \qquad\qquad
&&\big\{\dot Q_{\alpha\ad},\, \dot{\bar Q}_\alphad^\bd\big\}
=-\delta_\ad^\bd P_{\alpha\alphad}\,, \\
&\big\{S^{\alpha}_a ,\,\bar S^{\alphad b} \big\}
= \delta^b_a K^{\alpha\alphad}\,, \qquad\qquad
&&\big\{\dot S^{\alpha\ad} ,\,\dot{\bar S}^{\alphad}_\bd\big\}
=-\delta^\ad_\bd K^{\alpha\alphad}\,, \\
&\big[K^{\alpha\alphad},\, Q^a_{\beta} \big]=
\delta^{\alpha}_{\beta}\bar S^{\alphad a}\,, \qquad\qquad
&&\big[K^{\alpha\alphad},\, \dot{\bar Q}_\betad^\ad \big]
=\delta^{\alphad}_{\betad} \dot S^{\alpha\ad}\,, \\
&\big[P_{\alpha\alphad},\, S^{\beta}_a \big]
= -\delta^{\beta}_{\alpha}\bar Q_{\alphad a}\,, \qquad
&&\big[P_{\alpha\alphad},\, \dot{\bar S}^\betad_\ad\big]
=-\delta^{\betad}_{\alphad} \dot Q_{\alpha\ad}\,, \\
&\big\{ Q^a_{\alpha} ,\, S^{\beta}_b \big\}
= \delta^a_b J^{\beta}_{\,\,\alpha} +
\delta^{\beta}_{\alpha} R^a_{\ b}
+ \frac{1}{2} \delta^a_b \delta^{\beta}_{\alpha} (D+\dot D)\,, \qquad\ \
&&\big\{ Q^a_{\alpha} ,\, \dot S^{\beta\bd} \big\}
=\delta^{\beta}_{\alpha} \dot K^{\bd a}\,,\\
&\big\{\dot Q_{\ad\alpha} ,\, \dot S^{\beta\bd} \big\}
=-\delta_\ad^\bd J^{\beta}_{\,\,\alpha} 
+\delta^{\beta}_{\alpha}\dot R^{\bd}_{\ \ad}
-\frac{1}{2} \delta_\ad^\bd \delta^{\beta}_{\alpha} (D-\dot D)\,, \qquad
&&\big\{\dot Q_{\ad\alpha} ,\, S^{\beta}_b \big\}
=\delta^{\beta}_{\alpha} \dot P_{b\ad}\,,\\
&\big\{ \bar Q_{\alphad a} ,\, \bar S^{\betad b} \big\}
=-\delta^b_a \bar{J}^{\betad}_{\,\,\alphad} 
+\delta^{\betad}_{\alphad} R^b_{\ a}
-\frac{1}{2} \delta^b_a \delta^{\betad}_{\alphad}(D-\dot D)\,,\qquad
&&\big\{ \bar Q_{\alphad a} ,\, \dot{\bar S}^\betad_\bd \big\}
=-\delta_\alphad^\betad \dot P_{a\bd}\,,\\
&\big\{\dot{\bar Q}_\alphad^\ad ,\, \dot{\bar S}^\betad_\bd \big\}
=-\delta^\ad_\bd \bar{J}^{\betad}_{\,\,\alphad} 
-\delta^{\betad}_{\alphad} \dot R^\ad_{\ \bd}
-\frac{1}{2} \delta^\ad_\bd \delta^{\betad}_{\alphad} (D+\dot D)\,,\qquad
&&\big\{ \dot {\bar Q}_{\alphad}^\ad ,\, \bar S^{\betad b} \big\}
=-\delta_\alphad^\betad \dot K^{\ad b}\,.
\end{aligned}
\label{R4-commutators}
\end{equation}

The construction of the field $C$ in Section~\ref{sec-R4} involves an 
identification between the conformal group and the $R$-symmetry group. In 
particular this gives a canonical identification between the undotted 
indices of $SU(2)_L$ and $SU(2)_B$ and between the dotted ones of 
$SU(2)_R$ and $SU(2)_A$. This allows one to define the traced supersymmetry 
generators
\be
\begin{aligned}
Q&= Q^\alpha{}_\alpha = a^\dagger_\alpha\, c^\alpha\,,\qquad
&\dot{\bar Q}&=\dot{\bar Q}^\alphad{}_\alphad=b^\dagger_\alphad\, d^\alphad\,,\\
S&= S^\alpha{}_\alpha = c^\dagger_\alpha\, a^\alpha\,,\qquad
&\dot{\bar S}&=\dot{\bar S}^\alphad{}_\alphad = d^\dagger_\alphad\, b^\alphad\,.
\end{aligned}
\ee
These generators are invariant under the diagonal sums of 
the $SU(2)$ factors, but not over the full sum of the conformal group and 
$R$-symmetry group. The two generators that are invariant under that 
identification require fully tracing over the off-diagonal blocks in 
\eqn{broken-matrix}. The resulting two supercharges which anti-commute 
with each-other are
\be
\cQ^+=Q-\dot{\bar S}\,,\qquad
\cQ^-=\dot{\bar Q}-S\,.
\ee

Under this identification it is also possible to assign space-time indices 
to the $R$-symmetry generators and to the remaining supercharges. 
Using the usual $\gamma$ matrices (now with $a,\dot a$ indices) we have
\begin{equation}
\begin{aligned}
&\dot P_\mu=\dot P_{a\ad}\, (\bar \sigma_\mu)^{\ad a}
=R_{5\mu}+iR_{6\mu}\,, \qquad&
&\dot K_\mu=\dot K^{\ad a}\, (\sigma_\mu)_{a\ad}
=R_{5\mu}-iR_{6\mu}\,,\\
&R_{\mu\nu}=\frac{1}{2}\left (R^a{}_b\, (\sigma_{\mu\nu})_a{}^b
-\dot R^\ad{}_\bd\, (\bar\sigma_{\mu\nu})^\bd{}_\ad\, \right),\qquad&
&\dot D=iR_{56}\,.
\end{aligned}
\label{R4-rotations}
\end{equation}
For the supercharges we take the combinations
\be
\begin{aligned}
Q_\mu&=
(\bar\sigma_\mu)^{\ad\alpha}(\bar Q_{\alpha\ad}-\dot Q_{\alpha\ad})
\qquad\qquad
S_\mu=(\sigma_\mu)_{a\alphad}(\dot S^{\alphad a}-\bar S^{\alphad a})
\\
Q_{\mu\nu}&=\frac{1}{2}\left(
(S^a_\alpha-Q^a_\alpha)(\sigma_{\mu\nu})_a^{\ \alpha}
-(\dot{\bar Q}_\ad^\alphad-\dot{\bar S}_\ad^\alphad)
(\bar\sigma_{\mu\nu})_{\alphad}^{\ \ad}\right)
\\
Q_D&=\frac{1}{2}\left(
S^a_a-Q^a_a+\dot{\bar Q}_\ad^\ad-\dot{\bar S}_\ad^\ad\right)
\end{aligned}
\label{twisted-Qs}
\ee
Acting on them with $\cQ^\pm$ gives the twisted 
generators \eqn{R4-twisted} which are the sum of the 
conformal generators \eqn{conf-generators} and the $R$-symmetries 
\eqn{R4-rotations}
\be
\begin{aligned}
\big\{\cQ^\pm,\,Q_\mu\big\}
&=\hat P_\mu=P_\mu+\dot P_\mu\,,\qquad&
\big\{\cQ^\pm,\,Q_{\mu\nu}\big\}
&=\hat J_{\mu\nu}=J_{\mu\nu}+R_{\mu\nu}\,,\\
\big\{\cQ^\pm,\,S_\mu\big\}
&=\hat K_\mu=K_\mu+\dot K_\mu\,,\qquad&
\big\{\cQ^\pm,\,Q_D\big\}
&=\hat D=D+\dot D\,.
\end{aligned}
\label{twisted-comm-R4}
\ee
Under the action of these generators the field $C$ transforms as a 
dimension-zero scalar \eqn{twisted-C}.

We would like to comment that after choosing the scalar field $C$ \eqn{C}, it is 
natural to arrange the five other scalar fields as \cite{deMedeiros:2001kx}
\be
\begin{aligned}
V^\mu&=i\Phi^\mu+x^\mu(\Phi^6-i\Phi^5)\,,\\
B&=\Phi^6-i\Phi^5\,.
\end{aligned}
\ee
The full twisted conformal group \eqn{R4-twisted} as well as the 
twisted supercharges \eqn{twisted-Qs} give many more relations among 
the correlation functions of operators made of $C$ and these fields. For 
example the twisted conformal generators acting on $V^\mu$ give
\begin{equation}
\begin{aligned}
\hat P_\mu\,V^\nu&=\partial_\mu V^\nu\,,\\
\hat J_{\mu\nu}\,V^\rho
&=(x_\mu\partial_\nu-x_\nu\partial_\mu)V^\rho
+\delta_\mu^\rho V_\nu-\delta_\nu^\rho V_\mu\,,\\
\hat D\,V^\mu&=x^\nu\partial_\nu V^\mu+V^\mu\,,\\
\hat K_\mu\,V^\nu &=(2x_\mu x^\nu\partial_\nu+2x_\mu
-x^2\partial_\mu)V^\nu-2x_\mu V^\nu
+\delta^\nu_\mu(2x_\rho V^\rho-C)\,.
\end{aligned}
\label{twisted-V}
\end{equation}
and the action on $B$ is
\begin{equation}
\begin{aligned}
\hat P_\mu\,B&=\partial_\mu B\,,\\
\hat J_{\mu\nu}\,B
&=(x_\mu\partial_\nu-x_\nu\partial_\mu)B\,,\\
\hat D\,B&=x^\mu\partial_\mu B+2B\,,\\
\hat K_\mu\,B&=(2x_\mu x^\nu\partial_\nu+4x_\mu-x^2\partial_\mu)B-2V_\mu\,.
\end{aligned}
\label{twisted-B}
\end{equation}
We will not explore further the consequences of these relations here.

\section{Symmetry breaking for example II}
\label{app-R2-susy}

The construction of the field $C$ on $\bR^2$ involves choosing
three of the real scalars, so it explicitly breaks the $R$-symmetry
group $SU(4)\to SU(2)_{A'} \times SU(2)_{B'}$. Unlike the breaking in 
Appendix~\ref{app-R4-susy}, here the breaking is such that 
the $\bf 4$ of $SU(4)$ becomes the $({\bf2},\,{\bf2})$ of 
$SU(2)_{A'}\times SU(2)_{B'}$. Now the supercharges will carry indices 
of both groups, a dotted one for $SU(2)_{A'}$ and an undotted one for 
$SU(2)_{B'}$.

This breaking of $SU(4)\to SU(2)_{A'}\times SU(2)_{B'}$ is very similar to that 
required for the study of the supersymmetric Wilson loops of 
\cite{Drukker:2007dw,Drukker:2007qr} 
and much of this appendix is copied from Appendix~A of
\cite{Drukker:2007qr}.

The $R$-symmetry generators decompose under $SU(4) \rightarrow
SU(2)_{A'} \times SU(2)_{B'}$ as $\bf{15} \rightarrow (\bf{3},1) +
(1,\bf{3}) + (\bf{3},\bf{3})$. This can be explicitly written as
\begin{equation}
R^A_{\ B} \rightarrow R^{\ad a}_{\ \ \bd b} =
\frac{1}{2}\delta^a_b \dot T^{\ad}_{\ \bd} + \frac{1}{2}
\delta^{\ad}_{\bd} T^{a}_{\ b} + \frac{1}{2} M^{\ad a}_{\
\ \bd b} \label{SU4-break}
\end{equation}
where $\dot T^{\ad}_{\ \bd}$ and $T^{a}_{\  b}$ are
respectively the $SU(2)_{A'}$ and $SU(2)_{B'}$ generators, and the 9
generators in the $(\bf{3},\bf{3})$ are given by $M^{\ad a}_{\
\ \bd b}$, which is traceless in each pair of indices
\begin{equation}
\delta^{\bd}_{\ad} M^{\ad a}_{\ \ \bd b}=
\delta^{b}_{a} M^{\ad a}_{\ \ \bd b}=0\,.
\end{equation}

The commutation relations of the supercharges written in 
$SU(2)_{A'} \times SU(2)_{B'}$ notation are
\begin{equation}
\begin{aligned}
&\big\{ Q^{\ad a}_{\alpha},\, \bar Q_{\alphad\bd b}\big\}
=\delta^{\ad}_{\bd}\delta^a_bP_{\alpha\alphad}\,,
\qquad\,
\big\{ S^\alpha_{\ad a},\,\bar S^{\alphad \bd b} \big\}
=\delta_{\ad}^{\bd}\delta_a^b K^{\alpha\alphad}\,, \\
&\big[K^{\alpha\alphad},\, Q^{\ad a}_{\beta} \big]
=\delta^\alpha_\beta\bar S^{\alphad\ad a}\,, \qquad\quad
\big[K^{\alpha\alphad},\, \bar Q_{\betad \ad a}\big]
=\delta^{\alphad}_{\betad} S^\alpha_{\ad a} \,, \\
&\big[P_{\alpha\alphad},\, S^\beta_{\ad a}\big]
=\delta_\alpha^\beta \bar Q_{\alphad\ad a}\,, \qquad\quad~
\big[P_{\alpha\alphad},\, \bar S^{\betad \ad a}\big]
=\delta_{\alphad}^{\betad} Q_\alpha^{\ad a} \,, \\
&\big\{ Q^{\ad a}_{\alpha} ,\, S^\beta_{\bd b}\big\}
=\delta^{\ad}_{\bd}\delta^a_b J^\beta_{\ \alpha}+
\frac{1}{2}\delta_\alpha^\beta \left(\delta^a_b
\dot T^{\ad}_{\ \bd} + \delta^{\ad}_{\bd} T^a_{\ b}
+M^{\ad a}_{\ \ \bd b}+\delta^{\ad}_{\bd}\delta^a_b  D
\right)\,, \\
&\big\{ \bar Q_{\alphad \ad a},\,\bar S^{\betad\bd b}\big\}
=\delta_{\ad}^{\bd}\delta_a^b\bar{J}^{\betad}_{\ \alphad}
-\frac{1}{2}\delta_{\alphad}^{\betad} \left(
\delta_a^b\dot T^{\bd}_{\ \ad}+\delta_{\ad}^{\bd}T^b_{\ a}
+M^{\bd b}_{\ \ \ad a}-\delta_{\ad}^{\bd}\delta_a^bD
\right)\,.
\end{aligned}
\label{bigalgebra2}
\end{equation}

In Section~\ref{sec-R2} we use also the $R$-symmetry generators with 
$SO(6)$ vector indices $R_{ij}$. It is useful therefore to identify 
them, for $i=1,2,3$, with the rotations $\dot T^\ad_{\ \bd}$. 
This is done through
\begin{equation}
R_{ij}=-\frac{1}{2}(\rho_{ij})^\ad_{\ \bd}\,\dot T^{\bd}_{\ \ad}\,,
\qquad
\dot T^\ad_{\ \bd}=\frac{1}{2}(\rho^{ij})^\ad_{\ \bd}R_{ij}\,,
\qquad
(\rho_{ij})^\ad_{\ \bd}=i\,\varepsilon_{ijk}(\tau^k)^\ad_{\ \bd}\,.
\label{rhoij-choice}
\end{equation}
The commutators are
\begin{equation}
\big[\dot T^\ad_{\ \bd},\,\dot T^\cd_{\ \dot d}\big]
=-\delta^\ad_{\dot d}\,\dot T^\cd_{\ \bd}
+\delta^\cd_{\bd}\,\dot T^\ad_{\ \dot d}
\quad\Leftrightarrow\quad
\big[R_{ij},\,R_{kl}\big]=\delta_{ik}R_{jl}-\delta_{jk}R_{il}
+\delta_{il}R_{kj}-\delta_{jl}R_{ki}\,,
\end{equation}
like in (\ref{commute}).

\subsection{Action of $SL(2,\bR)\times SL(2,\bR)\times SU(2)$}
\label{app-R2-twist}

All operators in the plane transform in representations of the rigid conformal 
group $SL(2,\bC)\simeq SL(2,\bR)\times SL(2,\bR)$. The field $Z$ \eqn{Z} as 
well as $Y$ and $W$ \eqn{YW} carry also $SU(2)_{A'}$ indices and 
transform under this group. In Section~\ref{sec-R2-twist} we discussed 
the action of these generators, which we elaborate on here.

We write the holomorphic, anti-holomorphic and $SU(2)_{A'}$ algebras in terms 
of raising and lowering operators
\begin{align}
L_1&=\frac{1}{2}(P_1-iP_2)\,,\qquad&
L_0&=\frac{1}{2}(D-iJ_{12})\,,\qquad&
L_{-1}&=\frac{1}{2}(K_1+iK_2)\,,
\\
\bar{L}_1&=\frac{1}{2}(P_1+iP_2)\,,\qquad&
\bar{L}_0&=\frac{1}{2}(D+iJ_{12})\,,\qquad&
\bar{L}_{-1}&=\frac{1}{2}(K_1-iK_2)\,,
\\
R_+&=-i(R_{23}+iR_{31})\,,\qquad&
R_0&=iR_{12}\,,\qquad&
R_-&=i(R_{23}-iR_{31})\,,
\end{align}

The holomorpic operators act on the fields by
\begin{equation}
\begin{aligned}
L_1\,Z&=\partial_wZ\,,\qquad&
L_0\,Z&=w\,\partial_wZ+\frac{1}{2}Z\,,\qquad&
L_{-1}\,Z=w^2\,\partial_wZ+wZ\,,
\\
L_1\,Y&=\partial_wY\,,\qquad&
L_0\,Y&=w\,\partial_wY+\frac{1}{2}Y\,,\qquad&
L_{-1}\,Y=w^2\,\partial_wY+wY\,,
\\
L_1\,W&=\partial_wW\,,\qquad&
L_0\,W&=w\,\partial_wW+\frac{1}{2}W\,,\qquad&
L_{-1}\,W=w^2\,\partial_wW+wW\,.
\end{aligned}
\end{equation}
They all therefore transforms as a weight $1/2$ primary field of this
group. This is clearly the same behavior as for any of the scalar fields,
since there is no explicit $w$ dependence in
the definitions of $Z$, $Y$ and $W$

The action of the anti-holomorphic generators is more complicated
\be
\begin{aligned}
\bar{L}_1\,Z&=\partial_{\bar{w}}Z-2Y\,,\quad
&\bar{L}_0\,Z&=\bar{w}\,\partial_{\bar{w}}Z+\frac{1}{2}Z-2\bar wY\,,\\
\bar{L}_{-1}\,Z&=\bar{w}^2\,\partial_{\bar{w}}Z+\bar{w}Z-2\bar w^2Y\,,&&
\\
\bar{L}_1\,Y&=\partial_{\bar{w}}Y-W\,,\quad
&\bar{L}_0\,Y&=\bar{w}\,\partial_{\bar{w}}Y+\frac{1}{2}Y-\bar wW\,,\\
\bar{L}_{-1}\,Y&=\bar{w}^2\,\partial_{\bar{w}}Y+\bar{w}Y-\bar w^2W\,,&&
\\
\bar{L}_1\,W&=\partial_{\bar{w}}W\,,\quad
&\bar{L}_0\,W&=\bar{w}\,\partial_{\bar{w}}W+\frac{1}{2}W\,,\\
\bar{L}_{-1}\,W&=\bar{w}^2\,\partial_{\bar{w}}Z+\bar{w}W\,.&&
\end{aligned}
\label{R2-barL-action}
\ee
likewise for $SU(2)_{A'}$
\be
\begin{aligned}
R_+\,Z&=2Y\,,\quad
&R_0\,Z&=2\bar wY-Z\,,\quad
&R_-\,Z&=2\bar w^2Y-2\bar wZ\,,\\
R_+\,Y&=W\,,\quad
&R_0\,Y&=\bar wW\,,\quad
&R_-\,Y&=\bar w^2W-Z\,,\\
R_+\,W&=0\,,\quad
&R_0\,W&=W\,,\quad
&R_-\,W&=2\bar wW-2Y\,,
\end{aligned}
\label{R2-R-action}
\ee
and

The linear combination
\begin{equation}
\dot{L}_1=\bar{L}_1+R_+\,,\qquad
\dot{L}_0=\bar{L}_0+R_0\,,\qquad
\dot{L}_{-1}=\bar{L}_{-1}+R_-\,.
\end{equation}
Has a relatively simple action on the fields
\begin{align}
\dot{L}_1\,Z&=\partial_{\bar{w}}Z\,,\quad
&\dot{L}_0\,Z&=\bar{w}\,\partial_{\bar{w}}Z-\frac{1}{2}Z\,,\quad
&\dot{L}_{-1}\,Z&=\bar{w}^2\,\partial_{\bar{w}}Z-\bar{w}Z\,,
\label{R2-dotL-action}
\\
\dot{L}_1\,Y&=\partial_{\bar{w}}Y\,,\quad
&\dot{L}_0\,Y&=\bar{w}\,\partial_{\bar{w}}Y+\frac{1}{2}Y\,,\quad
&\dot{L}_{-1}\,Y&=\bar{w}^2\,\partial_{\bar{w}}Y+\bar{w}Y-Z\,,
\nn\\
\dot{L}_1\,W&=\partial_{\bar{w}}W\,,\quad
&\dot{L}_0\,W&=\bar{w}\,\partial_{\bar{w}}W+\frac{3}{2}W\,,\quad
&\dot{L}_{-1}\,W&=\bar{w}^2\,\partial_{\bar{w}}W+3\bar{w}W-2Y\,.
\nn
\end{align}

$Z$ therefore transforms as a weight $-1/2$ field of this
twisted anti-holomorphic $SL(2,\bR)$.
$Y$ has weight $1/2$ and $W$ has weight $3/2$, but they are
not primaries, as can be seen from the additional term in the action
of $\dot{L}_{-1}$.

A different combination of generators appears as the anti-commutator 
of the supercharges which annihilate $Z$ and the other supercharges. 
Those are
\begin{equation}
\hat{L}_1=\bar{L}_1+\frac{1}{2}R_+\,,\qquad
\hat{L}_0=\bar{L}_0+\frac{1}{2}R_0\,,\qquad
\hat{L}_{-1}=\bar{L}_{-1}+\frac{1}{2}R_-\,.
\end{equation}
They act on the fields by
\be
\begin{aligned}
\hat{L}_1\,Z&=\partial_{\bar{w}}Z-Y\,,\quad
&\hat{L}_0\,Z&=\bar{w}\,\partial_{\bar{w}}Z-\bar wY\,,\\
\hat{L}_{-1}\,Z&=\bar{w}^2\,\partial_{\bar{w}}Z-\bar w^2Y\,,
&&\\
\hat{L}_1\,Y&=\partial_{\bar{w}}Y-\frac{1}{2}W\,,\quad
&\hat{L}_0\,Y&=\bar{w}\,\partial_{\bar{w}}Y+\frac{1}{2}Y
-\frac{1}{2}\bar wW\,,\\
\hat{L}_{-1}\,Y&=\bar{w}^2\,\partial_{\bar{w}}Y+\bar{w}Y
-\frac{1}{2}\bar w^2W-\frac{1}{2}Z\,,
\hskip-1in&&\\
\hat{L}_1\,W&=\partial_{\bar{w}}W\,,\quad
&\hat{L}_0\,W&=\bar{w}\,\partial_{\bar{w}}W+W\,,\\
\hat{L}_{-1}\,W&=\bar{w}^2\,\partial_{\bar{w}}W+2\bar{w}W-Y\,.&&
\label{app-L-hat-action}
\end{aligned}
\ee

These generators indeed arise as the anti-commutators
\begin{align}
\big\{\cQ^+_a\,,iQ_2^{\ \dot1a}+\bar S^{\dot1\dot2a}\big\}
&=2\big(J^2_{\ 2}+\bar{J}^{\dot 1}_{\ \dot 1}+D\big)
+\dot T^{\dot1}_{\ \dot1}-\dot T^{\dot2}_{\ \dot2}
=2(D+iJ_{12}+iR_{12})
=4\hat L_0\,,
\nn\\
\big\{\cQ^+_a\,,-iQ_2^{\ \dot2a}\big\}
&=-2iP_{2\dot1}-\dot T^{\dot2}_{\ \dot1}
=P_1+iP_2-i(R_{23}+iR_{31})
=2\hat L_1\,,
\nn\\
\big\{\cQ^+_a\,,-\bar S^{\dot1\dot1a}\big\}
&=2iK^{\dot12}+\dot T^{\dot1}_{\ \dot2}
=K_1-iK_2+i(R_{23}-iR_{31})
=2\hat L_{-1}\,.
\end{align}
These expressions allow one to derive relations among correlation 
functions of operators made out of $Z$, $Y$ and $W$ as discussed at the
end of Section~\ref{sec-R2-twist}.

\section{Local operators on $S^2$}
\label{app-S2}

As was mentioned in Section~\ref{sec-R2}, there is also a natural definition 
for a scalar field coupling to three scalars on $S^2$. 
At the point $x^i\in S^2$ consider the following combination of
the three real scalar fields $\Phi^1$, $\Phi^2$ and $\Phi^3$
\eqn{Z-on-S^2}
\begin{equation}
Z^i=
(\delta^{ij}-x^ix^j)\Phi^j+i\varepsilon_{ijk}x^j\Phi^k\,,
\label{Zi}
\end{equation}
By virtue of the superscript, $Z^i$ is a three-dimensional vector. But
due to the identities
\begin{equation}
x^i\,Z^i=0\,,\qquad
\varepsilon_{ijk}\,x^j\,Z^k=-iZ^i\,,
\end{equation}
the three different components are related by a phase.

To deal with the ambiguity it is convenient to use complex coordinates
on $S^2$, through the stereographic projection
\begin{equation}
x^i=\frac{1}{1+w\bar{w}}
\left(w+\bar{w},\,-i(w-\bar{w}),\,1-w\bar{w}\right)\,.
\end{equation}
With this
\begin{equation}
Z^i=a^i\,\bar a^j\,\Phi^j\,,\qquad a^i=\frac{1}{1+w\bar{w}}
\left(-i(1-w^2),\,1+w^2,\,2iw\right)\,.
\end{equation}
So the index $i$ on $Z$ is related to the holomorphic coordinate
$w$, and we can eliminate it by defining
\begin{equation}
Z=\frac{1}{1+w\bar{w}} \left(i(1-\bar{w}^2)\Phi^1+(1+\bar
w^2)\Phi^2-2i\bar{w}\Phi^3\right)\,. \label{ZS2}
\end{equation}
This is exactly the same as (\ref{Z}), apart for a factor of
$(1+w\bar{w})$ due to the conformal transformation of the fields of
dimension one.

$Y$ and $W$ \eqn{YW} can also be defined as
\be
\begin{aligned}
2Y&=-i(x^1-ix^2)(\Phi^1+i\Phi^2)-(1+x^3)\Phi^3\,,\\
2W&=(1+x^3)(\Phi^2-i\Phi^1)\,.
\end{aligned}
\ee

\subsection{Supersymmetry}
By use of the stereographic projection, the operators made of these 
fields are analogous to those on the plane and any number of operators 
made of $Z$ \eqn{Zi} on the sphere will therefore share four supercharges. 

For completeness we perform the supersymmetry analysis also in this 
case. The supersymmetry variation of $Z$ gives
\begin{equation}
\delta Z\propto
\bar a^i\rho^i\left(\epsilon_0+x^j\gamma^j\epsilon_1\right).
\end{equation}
Expressing $\bar a^i$ and $x^j$ in terms of $w$ and $\bar{w}$ (or
alternatively working directly with the expression (\ref{Zi})) one
finds that the variation vanishes for arbitrary positions if
\begin{equation}
\rho^{12}\epsilon_0+i\gamma^3\epsilon_1=0\,,
\qquad
\rho^{23}\epsilon_0+i\gamma^1\epsilon_1=0\,,
\qquad
\rho^{31}\epsilon_0+i\gamma^2\epsilon_1=0\,.
\label{ep0-ep1}
\end{equation}
Eliminating $\epsilon_0$, we find the equations
\begin{equation}
\gamma^{12}\epsilon_1+\rho^{12}\epsilon_1=0\,,
\qquad
\gamma^{23}\epsilon_1+\rho^{23}\epsilon_1=0\,,
\qquad
\gamma^{31}\epsilon_1+\rho^{31}\epsilon_1=0\,.
\label{ep1-equations}
\end{equation}
This is the same as the condition for Wilson loops on $S^2$,
equation (2.23) in \cite{Drukker:2007qr}, up to an overall sign.
This means that there are solutions to these equations just like for
the Wilson loops, but the combined system of loops and local
operators is not supersymmetric.

To see exactly which supercharges annihilate our operators, consider
again the breaking of $SU(4)\to SU(2)_{A'}\times SU(2)_{B'}$ detailed in 
Appendix~\ref{app-R2-susy}. 
The combinations $\rho^{ij}$ act as Pauli matrices of $SU(2)_{A'}$. 
Likewise $\gamma^{ij}$ act as Pauli matrices on the
chiral and anti-chiral components of the spinors. For both chiralities
of $\epsilon_1$, which we label $\epsilon_1^\pm$ equation
(\ref{ep1-equations}) reads
\begin{equation}
(\tau^i_{L/R}+\tau^i_A)\epsilon_1^\pm=0\,,
\end{equation}
which means that $\epsilon_1^\pm$ is a singlet under the diagonal
group $SU(2)_{L/R}+SU(2)_{A'}$. Explicitly, using indices $\ad$ for
$SU(2)_{A'}$ and $a$ for $SU(2)_{B'}$ the solutions are given by the
two independent two-component spinors $\epsilon^{+\,a}$ and
$\epsilon^-_a$ as
\begin{equation}
\epsilon_{1\,\alpha}^{+\,\ad a}
=(\delta_\alpha^2\delta^\ad_{\dot1}
-\delta_\alpha^1\delta^\ad_{\dot2})\epsilon^{+\,a}
=i(\tau_2)^{\ad}_{\ \alpha}\epsilon^{+\,a}\,,
\qquad
\epsilon_{1\,\alphad\ad a}^-
=(\delta_\alphad^{\dot2}\delta_\ad^{\dot1}
-\delta_\alphad^{\dot1}\delta_\ad^{\dot2})\epsilon_a^-
=\varepsilon_{\alphad\ad}\epsilon_a^-\,.
\end{equation}

One then solves for $\epsilon_0$ using (\ref{ep0-ep1}). Note that
since these expressions have only a single $\gamma^i$ matrix, they
relate the $\epsilon_0$ and $\epsilon_1$ of opposite chiralities
\begin{equation}
\epsilon_0^{-\,\alphad\ad a}
=i(\delta^\alphad_{\dot1}\delta^\ad_{\dot2}
-\delta^\alphad_{\dot2}\delta^\ad_{\dot1})\epsilon^{+\,a}
=i\varepsilon^{\alphad\ad}\epsilon^{+\,a}\,,
\qquad
\epsilon^{+\,\alpha}_{0\,\ad a}
=i(\delta^\alpha_1\delta_\ad^{\dot2}
-\delta^\alpha_2\delta_\ad^{\dot1})\epsilon^-_a
=-(\tau_2)^\alpha_{\ \ad}\epsilon^-_a\,.
\end{equation}
Using all this (and remembering the signs in \eqn{general-susy-R2}) 
we can write the four supersymmetry generators as
\begin{equation}
\cQ_a=\bar Q_{\dot1\dot2a}-\bar Q_{\dot2\dot1a}-iS^2_{\ \dot1a}+iS^1_{\ \dot2a}
\,,\qquad
\dot\cQ^a=Q_1^{\ \dot2a}-Q_2^{\ \dot1a}+i\bar S^{\dot2\dot1a}
-i\bar S^{\dot1\dot2a}\,.
\end{equation}
The anti-commutator of the two gives
\begin{equation}
\big\{\cQ_a\,,\dot\cQ^b\big\}
=\delta_a^b(P_{1\dot1}+P_{2\dot2}+K^{\dot11}+K^{\dot22})+2iT^b_{\ a}
=\delta_a^b(P_4+K_4)+2iT^b_{\ a}\,.
\end{equation}
The trace is then the combination $P_4+K_4$ which maps the sphere at 
$x^4=0$ to itself and the second term is the $SU(2)_{B'}$ rotations, both 
are symmetries of all our operators $Z$ on $S^2$.

We can of course also consider all the other symmetry generators and their 
action on these fields. Again there are certain combinations of 
$SL(2,\bR)$ and $SU(2)$ generators with simple actions on these fields. 
These are completely analogous to what is detailed in 
Section~\ref{sec-R2-twist} and Appendix~\ref{app-R2-twist} and 
we do not repeat it.

\end{document}